\newif\ifsingle

\newif\ifFullVersion
\FullVersiontrue 

\ifsingle
\documentclass[11pt,draftclsnofoot, onecolumn]{IEEEtran}		
\else		
\documentclass[10pt,final, twocolumn]{IEEEtran}
\fi

\let\labelindent\relax  

\usepackage{times}
\usepackage{amsmath,dsfont}
\usepackage{amssymb}
\usepackage{epsfig,verbatim} 
\usepackage{setspace}
\usepackage{color}
\usepackage{cite}
\usepackage{epstopdf}
\usepackage{subfigure}
\usepackage{graphics}
\usepackage{accents}
\usepackage{acronym}
\usepackage[bookmarks,colorlinks]{hyperref}
\usepackage{mathtools}
\usepackage{algorithm}
\usepackage{algorithmic}
\usepackage{enumitem}
\usepackage[most]{tcolorbox}
\usepackage{cuted}

\newcommand{\myVec}[1]{{\boldsymbol{#1}}}

\newcommand{\mySet}[1]{\mathcal{#1}}
\newcommand{\SDGIter}{E}
\newcommand{\BatchSize}{B}
\newcommand{\NSelUsers}{K}
\newcommand{\Bandwidth}{b}
\newcommand{\ConvBound}{\Tilde{B}}
\newcommand{\Bits}{\beta}
\newcommand{\Npoints}{M} 
\newcommand{\LatFactor}{\bar{\sigma}_{\mathcal{L}}^2}
\newcommand{\ScaleFactor}{\zeta}
\newcommand{\DataInput}{\myVec{a}}
\newcommand{\DataLabel}{\myVec{b}}

\definecolor{mypurple}{rgb}{0.910, 0.910, 0.969}
\definecolor{myblue}{rgb}{0.122, 0.435, 0.698}

\acrodef{fl}[FL]{federated learning}
\acrodef{dnn}[DNN]{deep neural network}
\acrodef{iot}[IOT]{internet-of-things}
\acrodef{sgd}[SGD]{stochastic gradient descent}
\acrodef{mimo}[MIMO]{multiple-input multiple-output}

\ifFullVersion
\newcommand{\myCite}[1]{~\cite{#1}}
\else
\newcommand{\myCite}[1]{}
\fi

\newtheorem{theorem}{Theorem}
\newtheorem{lemma}{Lemma}

\newtheorem{definition}{Definition}

\newcommand{\beq}{\begin{equation}}
\newcommand{\eeq}{\end{equation}}
\newcommand{\bea}{\begin{array}}
\newcommand{\ena}{\end{array}}
\newcommand{\bds}{\begin {itemize}}
\newcommand{\eds}{\end {itemize}}
\newcommand{\bdf}{\begin{definition}}
\newcommand{\blm}{\begin{lemma}}
\newcommand{\edf}{\end{definition}}
\newcommand{\elm}{\end{lemma}}
\newcommand{\bthm}{\begin{theorem}}
\newcommand{\ethm}{\end{theorem}}
\newcommand{\bprp}{\begin{prop}}
\newcommand{\eprp}{\end{prop}}
\newcommand{\bcl}{\begin{claim}}
\newcommand{\ecl}{\end{claim}}
\newcommand{\bcr}{\begin{coro}}
\newcommand{\ecr}{\end{coro}}
\newcommand{\bquest}{\begin{question}}
\newcommand{\equest}{\end{question}}

\newcommand{\larrow}{{\larrow}}

\def\urltilda{\kern -.15em\lower .7ex\hbox{\~{}}\kern .04em}

\IEEEoverridecommandlockouts

\title{Federated Learning: A Signal Processing Perspective}
\author{
{Tomer Gafni, Nir Shlezinger, Kobi Cohen, Yonina C. Eldar, and H. Vincent Poor}
\thanks{
		T. Gafni, N. Shlezinger, and K. Cohen are with the School of Electrical and Computer Engineering, Ben-Gurion University of the Negev, Beer-Sheva, Israel (e-mail:gafnito@post.bgu.ac.il; \{nirshl, yakovsec\}@bgu.ac.il).
		 Y. C. Eldar is with the Math and CS Faculty, Weizmann Institute of Science, Rehovot, Israel (e-mail:  yonina.eldar@weizmann.ac.il). 
		H. V. Poor is with the ECE Department, Princeton University, Princeton, NJ  (e-mail:   poor@princeton.edu). 
	}
	\thanks{This research was partially supported by the ISRAEL SCIENCE FOUNDATION (grant No. 2640/20). 
}
}

\begin{document}

\maketitle
\pagestyle{plain}
\thispagestyle{plain}
	

\begin{abstract}
\label{sec:abstract}
The dramatic success of deep learning is largely due to the availability of data. Data samples are often acquired on edge devices, such as smart phones, vehicles and sensors, and in some cases cannot be shared due to privacy considerations. 
Federated learning is an emerging machine learning paradigm for training models across multiple  edge devices holding local datasets, without explicitly exchanging the data.
Learning in a federated manner differs from conventional centralized machine learning, and poses several core unique challenges and requirements, which are closely related to classical problems studied in the areas of signal processing and communications.
Consequently, dedicated schemes derived from these areas are expected to play an important role in the success of federated learning and the transition of deep learning from the domain of centralized servers to mobile edge devices.
In this article, we  provide a unified systematic framework for federated learning in a manner that encapsulates and highlights the main challenges that are natural to treat using signal processing tools.
We  present a  formulation for the federated learning paradigm from a signal processing perspective, and survey a set of candidate approaches for tackling its unique challenges. We further provide guidelines for the design and adaptation of signal processing and communication methods to facilitate federated learning at large scale.

\end{abstract}

\section{Introduction}
\label{sec:introduction}

%
Machine learning has led to breakthroughs in various fields, such as natural language processing, computer vision, and speech recognition. As machine learning methods, and particularly those based on \acp{dnn}, are data-driven, their success hinges on vast amounts of training data. This data is commonly generated at  edge devices, such as mobile phones, sensors, vehicles, and medical devices.  
In the traditional cloud-centric approach, data collected by mobile devices is uploaded and processed centrally at a cloud-based server or data center. As datasets, such as images and text messages, often contain private information, uploading them may be undesirable due to privacy and locality concerns. Furthermore, sharing massive datasets can result in a substantial burden on the communication links between the edge devices and the server.
These difficulties can be tackled by exploiting the available computational resources of edge devices via mobile edge computing, in order to carry out training at network edges without having to share the data \cite{chen2019deep}. 
An emerging paradigm for enabling learning at the edge is \textit{federated learning}, which features distributed learning  with centralized aggregations, orchestrated by an edge server (or multiple edge servers) \cite{mcmahan2017communication}. Federated learning is the focus of a growing research attention over the last few years \cite{kairouz2019advances}. 
In such systems, learning from distributed data and communicating between the  server and the devices are two critical and coupled aspects, and their fusion poses many new research challenges which are not encountered in traditional centralized deep learning \cite{li2020federated}. 

Federated learning is based on an iterative training process \cite{mcmahan2017communication}. At each federated learning iteration, the edge devices train a local model using their possibly private data, and transmit the updated model to the central server. The server aggregates the received updates into a single global model, and sends its parameters back to the edge devices. Therefore, to implement federated learning, edge devices only need to exchange their trained model parameters, avoiding the need to share their private data.
This property of federated learning makes it a promising solution for applications which are comprised of multiple entities that are required to learn from data, while operating under strict privacy practices~\cite{li2020review}.  
Federated learning, originally proposed by researchers from Google AI \cite{mcmahan2017communication}, has been applied to improve next-word prediction models in Google’s Gboard\myCite{hard2018federated}, allowing a multitude of smartphones to participate in training without requiring users to share their possibly private text messages. 
It has since been considered as an enabler technology for learning in applications with strict privacy considerations, such as models trained on healthcare data\myCite{xu2019federated}; highly distributed technologies, such as smart manufacturing systems\myCite{qu2020blockchained}; and learning over challenging communication networks, such as underwater\myCite{kwon2020multi} and unmanned aerial vehicle networks\myCite{brik2020federated}.  

%
Learning in a federated manner is subject to several key challenges that are not encountered in conventional cloud-centric learning \cite{li2020federated}.
These challenges include the communication bottleneck arising from the
need for a large number of users to repeatedly and  concurrently communicate their intermediate learned models, which are typically high dimensional, during the learning procedure; the heterogeneity of the participating devices in terms of storage, computation capabilities, and energy states, as well as the quantity and distribution of their local data; and privacy and security considerations following from, e.g., the possible presence of malicious users interfering with the training process.

The challenges encountered in federated learning affect the optimization performance and the accuracy of the learned model. They also introduce unique design considerations, such as communication burden and privacy guarantees, which arise from its distributed nature.
Moreover, federated learning adds potentially more hyperparameters to the optimization problem. These include separate tuning of the aggregation/global model update rule, number of edge devices selected per round, number of local steps per round, configuration of update compression algorithms, and more. These add to the existing numerous hyperparameters used in conventional deep learning, which, when combined with the aforementioned challenges, make the overall implementation and optimization of federated learning systems a difficult task. 

While the distributed and shared nature of federated learning involves characteristics which  are not encountered in conventional centralized deep learning, they are highly related to classical problems studied in the areas of signal processing and communications.
Consequently, the derivation of dedicated signal processing schemes is expected to play an important role in the success of federated learning. 
In fact, a multitude of federated learning-oriented methods based on established concepts in signal processing and communications have been recently proposed. These include the application of compression and quantization techniques to reduce the volume of the messages exchanged in the federated learning procedure \cite{lin2017deep,alistarh2017qsgd,shlezinger2020uveqfed,zheng2020design}; introducing functional and over-the-air computation tools  to facilitate high throughput federated aggregation over shared wireless channels \cite{zhu2019broadband,amiri2020federated, sery2020analog}; and the derivation of federated learning-aware resource allocation schemes to allow reliable communication between the participating entities during training \cite{chen2021communication,dinh2019federated}. 

In this article we present in a tutorial fashion the leading approaches for facilitating the implementation of federated learning at large scale using signal processing tools. 
As opposed to previous tutorials on federated learning, such as \cite{li2020federated}, which focused on the unique challenges of federated learning and its fundamental differences from centralized deep learning, here we focus on methods for tackling these challenges. 
To that aim, we discuss how the federated learning paradigm can be viewed from a signal processing perspective,  dividing its flow into three main steps: model distributing, local training, and global aggregation. We then focus on the global aggregation step, which involves conveying the local model updates from the users to the central server. We divide this step into three main phases which are carried out in a sequential fashion: $(a)$ encoding of the local model updates at the edge users into messages conveyed to the server; $(b)$ the transmission of the model updates and the allocation of the channel resources among the users, which take into account the statistical relationship between the input and the output of the physical communication channel; and $(c)$ combining (post-processing) at the server. 
For each stage, we elaborate on the specific aspects of federated learning which can benefit from tools derived in the signal processing and communication literature with proper adaptation.

We commence by reviewing the federated learning procedure, characterizing its goals and detailing the common federated averaging optimization algorithm. We then discuss the key challenges of federated learning, elaborating on how they arise from its three stage operation, and particularly from the global aggregation step, which captures the distributed operation of federated learning and its reliance on a multitude of diverse and possibly remote users. For each of the stages which comprise the global aggregation procedure, we  systematic identify its interplay with signal processing and communication methods by: $(a)$ reviewing the conventional treatment in the federated learning literature; $(b)$ identify how its procedure can be viewed from a signal processing perspective; $(c)$ discussing relevant tools based on this perspective and how they are applied in a non-federated-learning settings; $(d)$ identify the unique design considerations which one has to account for in order to adapt these tools for federated learning; and $(e)$ elaborate on existing methods from the recent literature along with detailed examples. The article concludes with a presentation of common guidelines for the derivation of future signal processing schemes for federated learning, along with an overview of some candidate issues for future research. 

\section{Basics in Federated Learning}
\label{sec:basics}
In this section we review the fundamentals of federated learning, identifying the unique role of signal processing in this emerging paradigm. We begin by describing the high level procedure by which federated learning enables distributed training of a centralized model. Then, we review conventional optimization mechanisms used in federated learning, after which we discuss the main challenges associated with the implementation of large scale federated learning systems. We conclude this section by formulating the three phases of federated learning, which are used in the division of our review of signal processing methods for federated learning in the following sections. 
%
%
\subsection{Federated Learning Flow}
\label{ssec:FL_flow}
Machine learning systems are data-driven, namely, their operation is not hard-coded, but learned from data. As such, machine learning  requires an {\em algorithm} that dictates what to learn from data as well as how to do so, i.e., the learning mechanism. For instance, deep learning relies on highly-expressive models based on \acp{dnn}, typically trained using variations of \ac{sgd} optimization; In addition, a sufficiently large {\em dataset} should be provided to the algorithm for learning purposes; Finally, adequate {\em computational resources} are necessary to process the data. In conventional centralized machine learning, all of these ingredients are available on the same system, which is typically a powerful cloud sever. However, in federated learning  these key ingredients are physically separated, as the data is divided among multiple entities, referred to as {\em edge users}.

\begin{figure}
    \centering
    \includegraphics[width=\columnwidth]{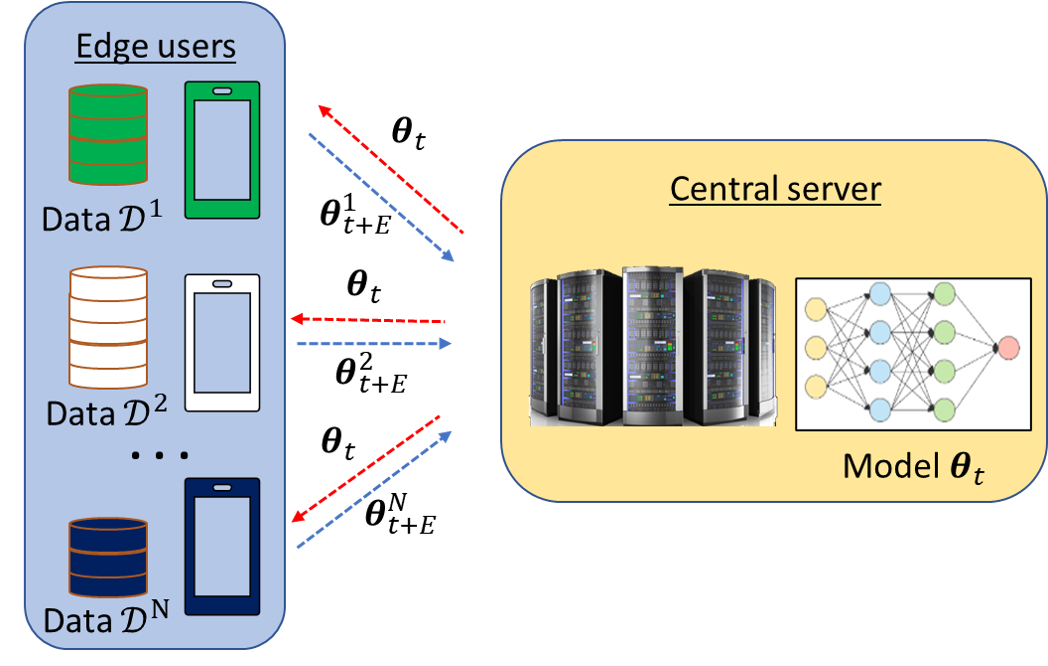}
    \caption{Federated learning setup illustration.}
    \label{fig:FL_HighLevel1}
\end{figure}

Federated learning deals with training of a single machine learning model, typically a \ac{dnn} maintained by a centralized server, without having the users share the data.  
While federated learning can be carried out with multiple servers, we focus our description on settings with a single server as illustrated in Fig.~\ref{fig:FL_HighLevel1}. As a result, the two main entities in federated learning systems are the edge users, which have access to local data, and the server that wishes to train a model in light of a given loss measure. 
We next formulate the objective and the operation of each of these entities, which are combined into the federated learning procedure:

\textit{Edge Users}: The data used to train the centralized model is collected by edge users, such as smartphones, wearable devices, and autonomous vehicles. The key differences of federated learning from conventional centralized learning are encapsulated in the properties of the edge devices. First, the edge users are more than just sensing devices; they possess sufficient computational resources needed to, e.g., locally train a \ac{dnn}. Furthermore, these devices are capable of interacting with the server and with each other over a communication network. Nonetheless, this communication is carried out over shared rate-limited channels, commonly the wireless cellular infrastructure. Finally, the users are not allowed to share their local data, due to, e.g., privacy constraints. Based on these properties, edge devices use their data to locally train a model, and share the updated model with the server.

To mathematically formulate the operation of the edge users, we consider a set of $N$ users, indexed by $\{1,\ldots,N\} \triangleq \mathcal{N}$. Each user of index $i \in \mathcal{N}$ has access to a local dataset $\mathcal{D}^i$. Focusing on a supervised setting, this dataset consists of $n_i$ labeled pairs $\mathcal{D}^i = \{\DataInput_n^i,\DataLabel_n^i\}_{n=1}^{n_i}$, where $\DataInput_n^i$ is referred to the training input, and $\DataLabel_n^i$ is the corresponding label. These datasets can be generated from different distributions, and are not necessarily independent and identically distributed (i.i.d.).

Each edge device of index $i$ uses its dataset $\mathcal{D}^i$ to train a \textit{local model} comprised of $d$ parameters, represented by the vector $\boldsymbol{\theta}^i \in \mathbb{R}^d$. Training is carried out  to minimize a local objective $f(\boldsymbol{\theta}; \mathcal{D}^i)$, based on a loss measure $\mathcal{L}(\cdot)$. The local objective for user $i$ is given by:
\begin{equation}
\label{eq:local_loss}
    \displaystyle f(\boldsymbol{\theta}; \{\DataInput_n^i,\DataLabel_n^i\}_{n=1}^{n_i}) = \frac{1}{n_i}\sum \limits_{n=1}^{n_i} \mathcal{L}(\DataInput_n^i,\DataLabel_n^i,\boldsymbol{\theta}).     
\end{equation}
Consequently, the objective of user $i$ is to recover the parameters $\boldsymbol{\theta}^i$ that minimize \eqref{eq:local_loss}, i.e.,
\begin{equation}
\label{eq:local_objective}
\displaystyle \boldsymbol{\theta}^{i^*} = \arg\min_{{\boldsymbol{\theta}}} f(\boldsymbol{\theta}; \mathcal{D}^i). 
\end{equation}

The fact that edge devices rely on data collected locally implies that each dataset $\mathcal{D}^i$ is typically comprised of a relatively small amount of samples, which may be insufficient to train an accurate machine learning model. Specifically, a model parametrized by $\boldsymbol{\theta}^{i^*} $ given in \eqref{eq:local_objective} is expected to yield relatively high loss values when applied to data samples which do not appear in $\mathcal{D}^i$, even if it is generated from the same distribution, i.e., the resulting model is likely to fail to generalize. Nonetheless, while each user has access to a limited amount of data, the number of users $N$ is typically very large in federated learning, motivating their collaboration in the learning procedure.

\textit{Centralized Server}:  The objective of the server is to utilize the data available at the users side to train a \textit{global model} with parameters $\boldsymbol{\theta}$. The objective used for learning $\boldsymbol{\theta}$ is given by:
\begin{equation}
\label{eq:global_loss}
    F(\boldsymbol{\theta}) = \sum \limits_{i=1}^N p_i \cdot f(\boldsymbol{\theta}; \mathcal{D}^i),
\end{equation}
where $p_i\geq 0$ are weighted average coefficients satisfying $\sum_{i=1}^N p_i = 1$, typically representing the portion of each user's dataset out of the overall training samples, i.e.,  $p_i = \frac{n_i}{\sum_{j=1}^N n_j}$ .
Consequently, the server aims to solve the following minimization problem:
\begin{equation}
\label{eq:global_objective}
    \displaystyle \boldsymbol{\theta}^* = \arg\min_{\boldsymbol{\theta}} F(\boldsymbol{\theta}).
\end{equation}

In principle, in order to recover $\boldsymbol{\theta}^*$, the server must have access to the complete dataset $\cup_{i=1}^{N}\mathcal{D}^i$, which is needed to compute the objective in \eqref{eq:global_loss}. Federated learning  allows users to collaboratively train a global model while keeping personal data on their devices, in a manner which is orchestrated by the centralized server, as detailed next.

\textit{Federated Learning Procedure}: In order to train the global model, federated learning iteratively combines local training based on \eqref{eq:local_objective} with centralized aggregation to approach the desired global model in \eqref{eq:global_objective}. In particular, each round of index $t$ is comprised of three main steps:
\begin{enumerate}
    \item Model distributing: The server  sends the updated global model $\boldsymbol{\theta}_{t}$ to the users, and each user sets its current local model to be $\boldsymbol{\theta}_{t}^i = \boldsymbol{\theta}_{t}$. 
    \item Local training: Each user $i$ which participates in the training procedure uses its local data to train the local model $\boldsymbol{\theta}_t^i$ into an updated model $\boldsymbol{\theta}_{t+1}^i$ according to \eqref{eq:local_objective} via a local optimization algorithm (e.g., several \ac{sgd} steps).
    \item Global aggregation: The participating users convey their local updates to the server, which processes the updated local models $\{\boldsymbol{\theta}^i_{t+1}\}$ into the global model $\boldsymbol{\theta}_{t+1}$.
\end{enumerate}
Steps (1)-(3) are repeated iteratively until convergence. While our description considers supervised learning, the same flow also applies for unsupervised settings, where the difference is that the loss measure in \eqref{eq:local_loss} is computed based solely on input samples $\{\myVec{a}_n^i\}_{n=1}^{n_i}$.

The above steps describe the main federated learning flow, which incorporates various different algorithms \cite{kairouz2019advances}. These methods specify how each of these steps is carried out. The most common mechanism adopted by the majority of federated learning schemes is the {\em Federated Averaging (FedAvg)} learning method \cite{mcmahan2017communication}, detailed in the following subsection.

%
%
\subsection{Federated Averaging}
\label{ssec:Learning_Algorithms}
\textit{FedAvg} specializes the federated learning procedure detailed in the previous subsection by setting the aggregation mapping to implement weighted averaging with the weights $\{p_i\}$ in \eqref{eq:global_loss}.

To mathematically formulate this learning scheme, we focus here on the implementation of \textit{FedAvg} where the users carry out their local optimization using the \ac{sgd} algorithm, referred to as \textit{local \ac{sgd}} \cite{li2019convergence}. Each user carries out $\SDGIter>0$ \ac{sgd} iterations with mini-batch size $\BatchSize$ prior to their global aggregation and distribution, where each such update is given by
\begin{equation}
    \label{eq:SGD}
     \boldsymbol{\theta}_{t+1}^i = \boldsymbol{\theta}_t^i - \eta_t \nabla f(\boldsymbol{\theta}_t^i; \mySet{D}_t^i).
\end{equation}
Here, $\eta_t >0$ is the step-size, also referred to as the learning rate;  $\mySet{D}_t^i$ is a randomly sampled mini-batch with $|\mySet{D}_t^i| = B$; and $\nabla f(\boldsymbol{\theta}_t,\mySet{D}_t^i) \in \mathbb{R}^d$ is the stochastic gradient of the objective function with respect to the model parameters evaluated at $\boldsymbol{\theta}_t$. 

After $\SDGIter$ \ac{sgd} iterations of the form \eqref{eq:SGD}, i.e., at iteration $t=k\SDGIter$ where $k$ is a positive integer, the set of users which participate in the current round, indexed by the set $\mySet{G}_{t} \subseteq \mySet{N}$, convey their updated models to the server. Typically, the  users convey the updates to the model generated in the current round, i.e., $\boldsymbol{\theta}_{k\SDGIter}^i - \boldsymbol{\theta}_{(k-1)\SDGIter}^i$, rather than the model weights $\boldsymbol{\theta}_{k\SDGIter}^i$. As the server knows $\boldsymbol{\theta}_{(k-1)\SDGIter}^i$, it can recover $\boldsymbol{\theta}_{k\SDGIter}^i$ from the difference $\boldsymbol{\theta}_{k\SDGIter}^i - \boldsymbol{\theta}_{(k-1)\SDGIter}^i$, while the latter tends to become sparse as convergence is approached.  
The server then aggregates these local updated models into a new global model $\boldsymbol{\theta}_{t+1}$, where $t=k\SDGIter$, by computing a weighted average
via \cite{li2019convergence}\footnote{Alternative averaging-based aggregation mappings have also been considered in FedAvg, such as the inclusion of the previous model of non-participating users in the averaging \cite{mcmahan2017communication}. For consistency, we refer to FedAvg as federated learning with server-side averaging via \eqref{eq:FedAvg_Server}, as in \cite{li2019convergence}.}:
\begin{equation}
\label{eq:FedAvg_Server}
    \boldsymbol{\theta}_{t+1} = \frac{N}{|\mathcal{G}_t|} \sum \limits_{i \in \mathcal{G}_t} p_i\boldsymbol{\theta}^i_{t+1}.
\end{equation}

The new global model is broadcast back  the users, which synchronize their local model accordingly.
The updating rule of the local SGD algorithm is thus given by:
\begin{equation}
\label{eq:FedAvg}
        \displaystyle \boldsymbol{\theta}_{t+1}^i \!=\! 
    \begin{cases}
    \boldsymbol{\theta}_t^i - \eta_t \nabla f(\boldsymbol{\theta}_t^i; \mySet{D}_t^i), & \text{if} \; t \notin \mathcal{I}_T,    \\ 
    \frac{N}{|\mySet{G}_t|} \sum \limits_{j \in \mySet{G}_t}\!p_j \left(\boldsymbol{\theta}_t^j \!- \!\eta_t \nabla f(\boldsymbol{\theta}_t^j; \mySet{D}_t^j)\right), & \text{if} \; t \in \mathcal{I}_T.
    \end{cases} 
    \end{equation}
    Here, $T$ is the number of total steps, and $\mathcal{I}_T \subseteq \{1,\ldots, T\}$ denotes a set of synchronization indices, i.e., the set of integer multiples of $\SDGIter$. 
If $\mathcal{I}_T = \{1,\ldots, T\}$, then $\SDGIter = 1$ and the aggregation is performed every iteration. If $\mathcal{I}_T = \{ T \}$, then aggregation only happens at the end, i.e., $\SDGIter = T$, which is known as one-shot averaging. This iterative procedure converges to the desired optimal global model \eqref{eq:global_objective} for strongly-convex and smooth objectives with bounded gradients (see \emph{Convergence of Local SGD} on Page \pageref{Box:FedAvg_convergence}) in the same asymptotic rate as mini-batch \ac{sgd} achieves when carried out in a centralized manner, i.e., when the server has access to the full dataset $\cup_{i=1}^{N}\mathcal{D}^i$. 

%

\begin{tcolorbox}[float*=t,
    width=2\linewidth,
	toprule = 0mm,
	bottomrule = 0mm,
	leftrule = 0mm,
	rightrule = 0mm,
	arc = 0mm,
	colframe = myblue,
	colback = mypurple,
	fonttitle = \sffamily\bfseries\large,
	title = Convergence of Local SGD]	
The unique challenges associated with federated learning affect the convergence analysis of the optimization algorithms, and add more hyper-parameters. 
Here we present the analysis for the convergence rate of the \textit{FedAvg} algorithm with local SGD which is described in (\ref{eq:FedAvg}), derived from \cite{li2019convergence}. This bound will serve as a benchmark when discussing the performance of algorithms based on signal processing methods throughout the tutorial.

It is assumed that the local objective functions $\{f(\boldsymbol{\theta};\mySet{D}^i)\}_{i\in\mathcal{N}}$ are all $L$-smooth, $\mu$-strongly convex, the variance of stochastic gradients in each device is bounded by $\sigma^2_i$, the expected square norm of the stochastic gradients is uniformly bounded by the parameter $G^2$, and $\Gamma$ is defined as the degree of heterogeneity between users. The \textit{FedAvg} algorithm terminates after $T$ iterations and returns $\boldsymbol{\theta}_T$ as the solution.   
Choosing $\gamma = \max\{8 \frac{L}{\mu},E\}$ and learning rate $\eta_t = \frac{2}{\mu(\gamma + t)}$, \textit{FedAvg} with full device participation satisfies 
\begin{equation}
    \label{eq:FedAvg_Convergence}
\displaystyle \mathbb{E}[F(\boldsymbol{\theta}_T)] - F^* \leq \frac{L}{\gamma+T-1}\bigg(\frac{2\ConvBound}{\mu^2} + \frac{\gamma}{2} \mathbb{E}||\boldsymbol{\theta}_0 - \boldsymbol{\theta}^* ||^2\bigg), 
\end{equation}
where $
    \displaystyle \ConvBound = \sum \limits_{i=1}^N p_i^2 \sigma_i^2 + 6L\Gamma + 8(E-1)^2G^2$. 
    
The main insight is that \textit{FedAvg} converges to the global optimum at a rate of $\mathcal{O}(1/T)$, which is the same asymptotic rate as mini-batch \ac{sgd} achieves when carried out in a centralized manner.    
The parameter $E$ (number of local SGD iterations prior to the aggregation) controls the convergence rate: Setting $E$ to be very small may lead to high communication cost. On the other hand, if $E$ is large, then $\boldsymbol{\theta}_t^i$ can converge to the minimizer of $f(\boldsymbol{\theta};\mySet{D}^i)$. 
It is noted that \textit{FedAvg} requires diminishing learning rates for convergence.  
\label{Box:FedAvg_convergence}
\end{tcolorbox}

%
%
\subsection{Challenges}
\label{ssec:challenges}
\begin{figure*}
    \centering
    \includegraphics[width=0.9\linewidth]{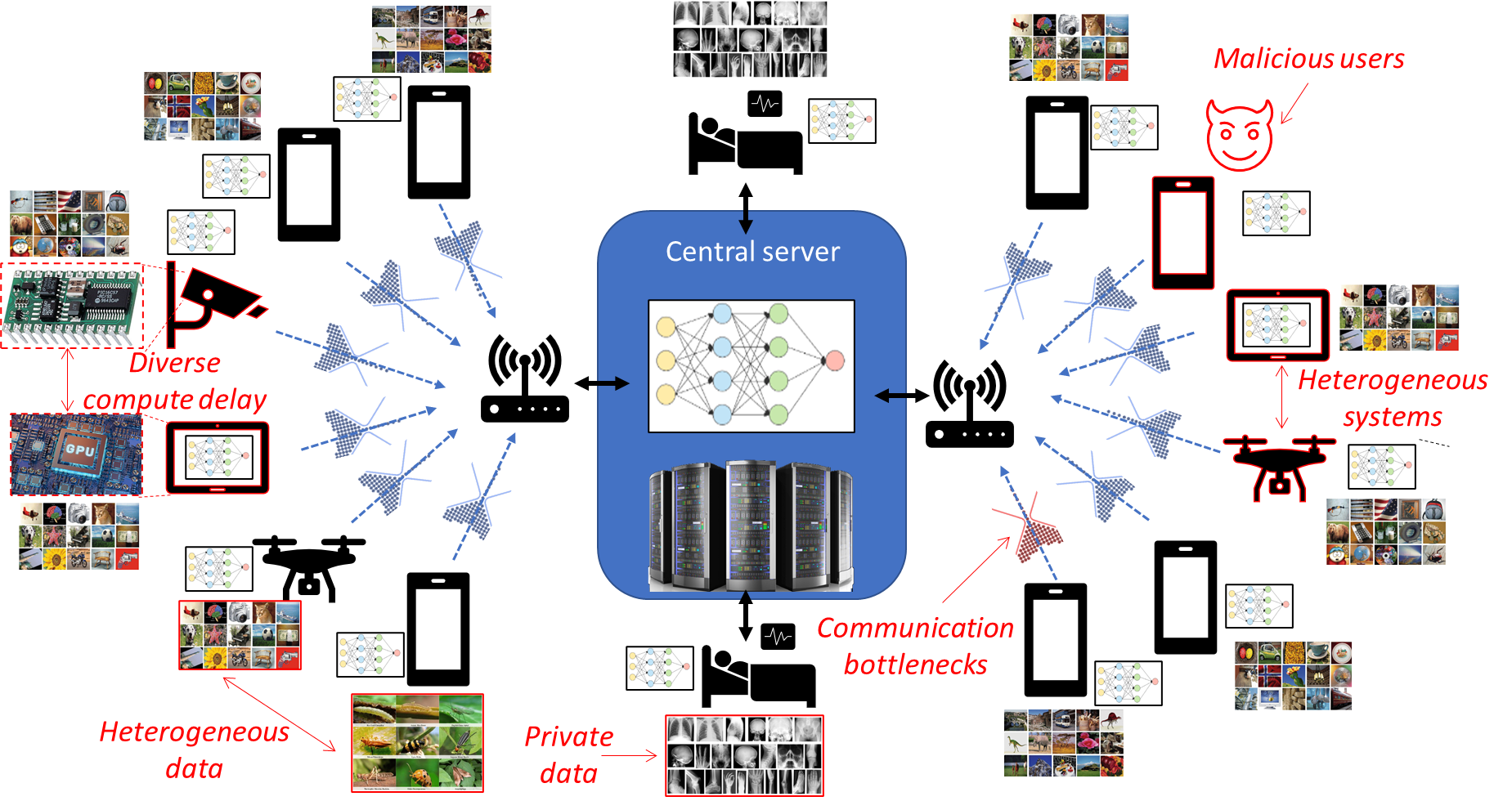}
    \caption{A federated learning system with various diverse users, highlighting some of the core challenges associated with federated learning, including communication bottlenecks; statistical heterogeneity of the data; heterogeneity in the different user devices; diverse hardware and computational delays; the need to preserve privacy with respect to the data; and the possible presence of malicious users.}
    \label{fig:Constraints}
\end{figure*}

Learning in a federated manner differs from conventional centralized deep learning, where the complete dataset is available to the server. This leads to several core challenges which are not encountered in conventional deep learning. These challenges, illustrated in Fig.~\ref{fig:Constraints},  are discussed in the following.

\subsubsection{\textbf{Communication Bottleneck}}
\label{sssec:communication_bottelneck}
The repeated exchange of updated models between the users and the server often involves massive transmissions over rate-limited communication channels. This challenge is particularly relevant when the model being trained is a \ac{dnn} comprised of a large amount of trainable parameters, as well as for federated learning carried out over shared and resource-limited wireless networks, e.g., when the users are wireless edge devices.  
In addition to overloading the communication infrastructure, these repeated transmissions imply that the time required to tune the global model depends not only on the number of training iterations, but also  on the delay induced by transmitting the model updates at each federated learning iteration\myCite{chen2020wireless}. 
Federated learning networks are potentially comprised of a massive number of energy-constrained edge users communicating over the wireless media. Considering the limited capacity of wireless channels, the communications of the model updates can be slower than local computations by  orders of magnitude. Hence, the communication bottleneck directly affects the training time of global models trained in a federated manner, which in turn may degrade their resulting accuracy. 

\subsubsection{\textbf{Statistical Heterogeneity}}
\label{sssec:statistical_heterogeneity}
Statistical heterogeneity implies that the data generating distributions vary between different sets of users. This is typically the case in federated learning, as the data available at each user device is likely to be personalized towards the specific user. 
Statistical heterogeneity implies that when training several instances of a model on multiple edge devices, each instance may be biased. As a result, the conventional strategy based on averaging the trained updates may not reflect the intended usage of the global model in inference.

\subsubsection{\textbf{System Heterogeneity}}
\label{sssec:system_heterogeneity} 
As federated learning involves a multitude of user devices, their behavior and availability can be quite diverse. This form of heterogeneity, referred to as {\em  behavior heterogeneity},  can affect the learning process that relies on the device status. For instance, one may design the federated learning mechanism to involve portable user devices only when they are either idle, charging, or connected to an unmetered network, such as Wi-Fi. As a result, participants may be unreliable and can drop out at any time, leading to uneven participation over the training group. 
Moreover, as opposed to centralized learning where training is carried out using a computationally powerful server, the participating devices in federated learning vary in terms of hardware specifications like computation and energy resources. For instance, a federated learning network may involve user devices ranging from sophisticated autonomous vehicles to hardware-limited \acl{iot} devices. This implies that the duration of the local training procedure can vary considerably between users, resulting in computation bottlenecks and straggler effects which delay the overall convergence of the learned model.

\subsubsection{\textbf{Privacy and Security}}
\label{sssec:privacy_security}
One of the key motivations for federated learning is to preserve privacy with respect to the users' data. In that sense, federated learning has distinct privacy advantages compared to centralized training on persisted data, as the information conveyed from the users is not the data itself but the model updates obtained using this local data.
However, one must also consider that information can be inferred from the learning process and  be traced back to its source in the resulting trained model. For instance, gradients are known to be leaky, and may  reveal the data used for computing them \cite{zhu19deep}. 
Furthermore, since the users are outside of the control of the server they may be unreliable, or even malicious, giving rise to security issues. Such users send the server arbitrary or tainted model updates, affecting the trained model, while exploiting the interactiveness of the process to have the opportunity to adapt.

%
\subsection{Signal Processing Stages in Federated Learning}
\label{ssec:SP_stages}
As detailed above, the federated learning procedure is comprised of three main steps: model distributing, local training, and global aggregation. Nonetheless, the unique challenges of federated learning discussed in the previous subsection are mostly associated with the last step of global aggregation. In particular, the distribution step involves the broadcasting of the global model from the server to the users, and thus is far less affected by privacy, security and heterogeneity considerations, compared to the transmission from the users to the server. Furthermore, this broadcasting is carried out over the downlink channel, which is typically less limited in terms of throughput compared to the uplink channel, and thus is less affected by communication limitations compared to the aggregation stage. 
The local training step, which yields the updated model based on the previous model and the data, is typically carried out using conventional optimizers, such as \ac{sgd} and its variants. Consequently, the federated learning step in which most challenges arise is during global aggregation. 

\begin{figure*}
    \centering
    \includegraphics[width=0.9\linewidth]{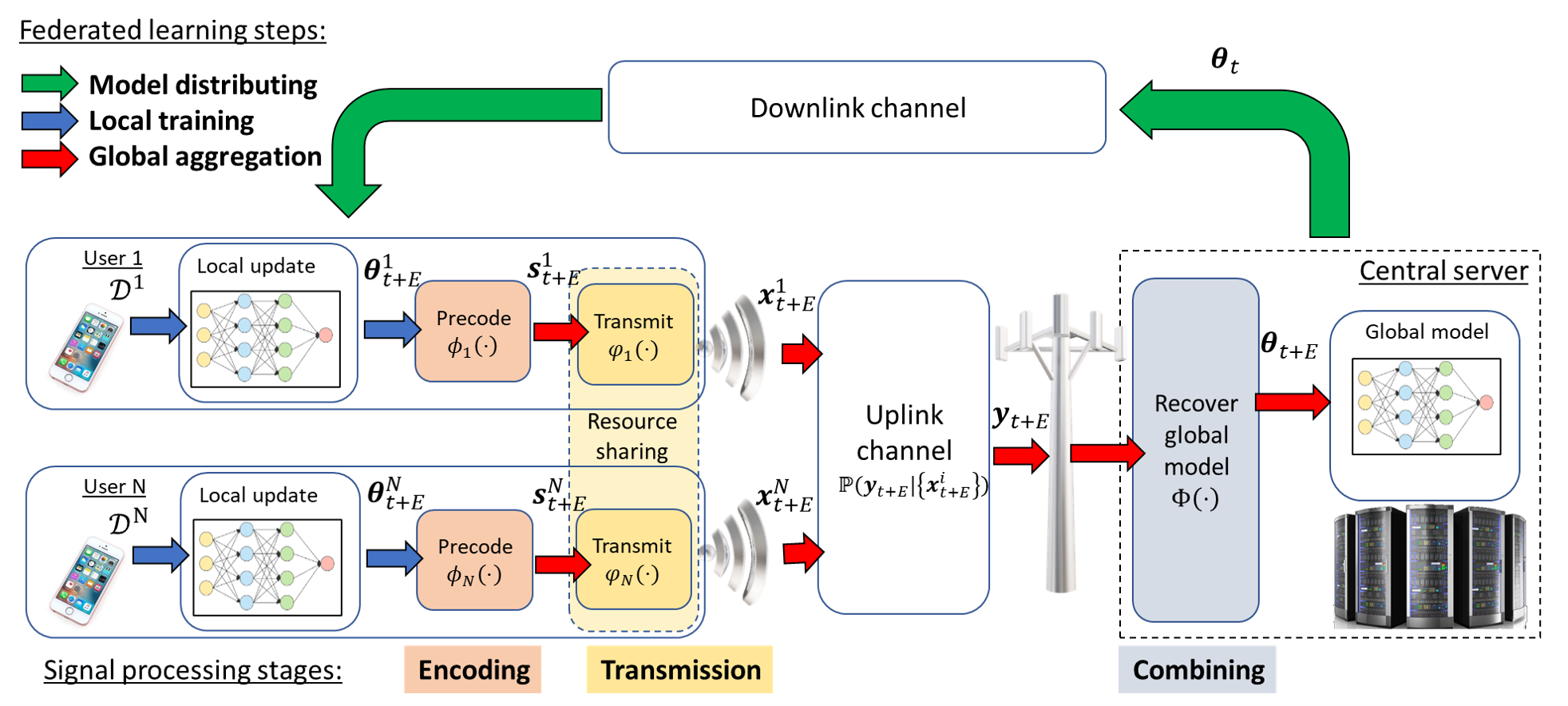}
    \caption{The federated learning procedure divided into the three main steps of model distributing, local training, and global aggregation. The latter is further represented as a three-stage process, involving the encoding of the model updates; their transmission to the server; and their combining into a global model.}
    \label{fig:FL_Flow1}
\end{figure*}

The global aggregation can be divided into three main stages, where each can be treated as a signal processing and/or communication task. These stages are: $1)$ the processing and encoding of the local training outcome at the edge users; $2)$ the transmission of these outcomes over shared wireless channels; and $3)$ the processing and combining of the received signals at the server as part of the global learning procedure. An illustration of the federated learning procedure including this division into stages is illustrated in Fig.~\ref{fig:FL_Flow1}.

%
\subsubsection{\textbf{Local Updates Processing and Encoding}} 
The first stage focuses on processing the local outcomes of the training procedure at the users' side before transmission. 
Such processing is essential to be able to convey the model updates to the server over a rate-limited shared link in a reliable and privacy preserving manner.
Given the local model for user $i$ at time $t$, $\boldsymbol{{\theta}}_t^i$, the local processing is the mapping:
\begin{equation}
\label{eq:precoding}
    \boldsymbol{s}_t^i = \phi_i(\boldsymbol{{\theta}}_t^i),
\end{equation}
where $\boldsymbol{s}_t^i$ is the processed model update, used to form the channel input in the transmission stage, as illustrated in Fig.~\ref{fig:FL_Flow1}.

The mapping \eqref{eq:precoding} should be designed to address the necessity to encode and compress the model updates. As we elaborate in the ``\nameref{sec:encoding}" section, this can be achieved by setting $\phi_i(\cdot)$ to implement stochastic quantization and sparsification of the model updates for the reduction of the communication load. 
Furthermore,  $\phi_i(\cdot)$ can be designed to boost privacy preservation with respect to the data, via, e.g., inducing local noise perturbations relying on the algorithmic foundations of differential privacy. 

%
\subsubsection{\textbf{Uplink Transmission}} 
The processed updates $\boldsymbol{s}_t^i$ are transmitted to the server for global updating, typically over a wireless channel. To that aim, each user must form its corresponding channel input, denoted $\boldsymbol{x}_t^i$, in order to convey $\boldsymbol{s}_t^i$, via the mapping
\begin{equation}
    \label{eq:transmission}
    \boldsymbol{x}_t^i = \varphi_i\left( \boldsymbol{s}_t^i\right). 
\end{equation}
The statistical relationship between the channel input $\boldsymbol{x}_t^i$ and the output obtained at the server side $\boldsymbol{y}_t$, is determined by the channel characteristics encapsulated by the conditional distribution $\mathbb{P}(\boldsymbol{y_t}| \{\boldsymbol{x}_t^i\}_i)$, as illustrated in Fig.~\ref{fig:FL_Flow1}.
That is, the server receives a noisy version of the channel inputs transmitted by the users, where the transmission mapping and the characteristics of the channel dictate the relationship between the channel output $\boldsymbol{y}_t$ observed by the server and the encoded model updates $\boldsymbol{s}_t^i$. 

As discussed in the ``\nameref{sec:Transmission}" section, methods related to the transmission phase typically focus on federated learning carried out over shared wireless channels. Here, the aim to allow high throughput and reliable communication in a manner which does not introduce notable delays to the overall learning procedure. 
This requires consideration of two main aspects: The first is the allocation of the resources of the channel in terms of bandwidth and transmission time. 
A special case of channel resources division is user selection and scheduling, determining the set of participating users in the current round $\mathcal{G}_t$. 
Furthermore,  one can exploit the shared nature of the wireless media as a form of over-the-air functional computation, allowing the users to exploit its full resources while benefiting from the resulting interference.   

%
\subsubsection{\textbf{Global Combining}} 
The server uses  its observed channel output $\boldsymbol{y}_t$, which contains the information on $\{\boldsymbol{\theta}_t^i\}_{i=1}^N$, in order to update the global model. This operation, carried out by the server, can be represented by the following mapping:
\begin{equation}
\label{eq:post_processing}
    \boldsymbol{{\theta}}_t = \Phi(\boldsymbol{y}_t).     
\end{equation}
Here, the server uses the channel output, which contains information regarding the individual model updates, in order to estimate a combined global model as illustrated in Fig.~\ref{fig:FL_Flow1}.  As we elaborate in the ``\nameref{sec:combining}" section, this mapping can also be utilized to compensate for noise and fading induced by the channel; improve the accuracy of the global model during inference by mitigating the effect of statistical heterogeneity; and overcome the possible harmful effect of malicious devices using Byzantine-robust aggregation.

\section{Local Updates Processing and Encoding}
\label{sec:encoding}
The first stage in the global aggregation step of federated learning focuses on the processing of the local outcomes of the training procedure at the users' side before their transmission.
%
Here, we  describe conventional encoding strategies, and  discuss how this operation can be viewed from a signal processing perspective, briefly reviewing related methods applied in a non-learning setting. We then discuss their adaptation to  encoding the local updates in federated learning by highlighting the specific design considerations and reviewing some candidate techniques from the recent literature.    

Given the local model for user $i$ at time instance $t$, $\boldsymbol{\theta}_t^i$, the local encoding process is given by the mapping $\boldsymbol{s}_t^i = \phi_i(\boldsymbol{{\theta}}_t^i)$ in  \eqref{eq:precoding}. 
As the server is interested in aggregating the model $\boldsymbol{\theta}_t^i$ and not in the encoded symbols $\boldsymbol{s}_t^i$, the encoding procedure should be combined with an appropriate joint decoding procedure.   
Since the local data sets possibly differ from each other in their size and distribution, each user will in general have an individual  mapping $\phi_i(\cdot)$, induced from its unique characteristics and based on the underlying task.

\textbf{Conventional Encoding}: 
The straight-forward implementation of the FedAvg algorithm requires the users to send the model updates to the server. In this case, the codeword is merely the difference between  the last distributed global model, which is available to the server and is identical among all users, and the current locally trained one, i.e., 
\begin{equation}
\label{eqn:ModelUpdate}
    \phi_i(\boldsymbol{{\theta}}_t^i) = \boldsymbol{\theta}_t^i - \boldsymbol{\theta}_{t-\SDGIter}^i.
\end{equation}

This conventional encoding does not account for any knowledge one may possess about the model updates and their structure. 
Furthermore, the full model update vector in \eqref{eqn:ModelUpdate} is of the same dimensionality as the model itself, implying that communicating it to the server is likely to induce a notable communication bottleneck. 
In addition, one of the main motivating factors behind federated learning is that the local data at each user does not leave the local device, as only gradients or the model updates computations from each user are shared with the server. However, even exchanging gradients in a raw form can leak information \cite{zhu19deep}.
Finally, \eqref{eqn:ModelUpdate} is  agnostic of the federated learning  task, as the server averages the updates from the users, and this aggregation procedure can mitigate noise on the updates induced by encoding techniques.  


\textbf{Signal Processing Perspective}: 
The mapping $\boldsymbol{\theta}_t^i \mapsto \boldsymbol{s}_t^i$, carried out at multiple user devices on each round, is in fact a form of {\em distributed source coding} \cite{xiong2004distributed}. The main objective in the source coding literature is compression, i.e., the goal is to map $\boldsymbol{\theta}_t^i$ into a compact representation from which the original vector can be recovered with minimal errors. Federated learning is also concerned with recovering a unified accurate model from these representations, typically via averaging, as well as preserve privacy with respect to the local datasets.  
This gives a rise to learning-aware encoding schemes, which can be divided into uplink compression and privacy preservation, as detailed in the following.
%

%
\subsection{Uplink Compression}
\label{ssec:uplink_compression}
Uplink compression is the procedure of encoding the model updates, having the codeword conveyed to the server being comprised of fewer (and preferably much fewer) bits compared to those used for representing the original model. 
Compression is necessary in order to reduce the uplink communication cost, thus addressing the communication bottleneck which is among the core challenges of federated learning.  

\textbf{Source Coding for Compression}:
Compression is the objective of traditional source coding methods. 
Source coding techniques are typically divided into lossless and lossy compression. Lossless compression, such as entropy coding, converts discrete quantities into a reduced-bit representation in a reversible manner. Lossy compression, such as quantization, can be applied also to continuous-valued quantities, and yields a compact digital representation from which the original input can be recovered only up to some error. As model parameters trained in a federated manner, e.g., the weights of \acp{dnn}, are typically treated as continuous-valued quantities during training, we henceforth focus on lossy compression methods. Furthermore, lossy compression allows achieving improved compression rates compared to lossless compression for discrete-valued weights (though at the cost of possible distortion upon recovery). Finally, it is common to combine lossy and lossless compression, e.g., first quantize and then apply entropy coding, as in entropy-constrained quantization. 

Source coding techniques also extend to distributed setups, where multiple users encode correlated local sources (e.g., vectors) such that the resulting representations are both compact in terms of number of bits, and can be jointly recovered using a single decoder \cite{xiong2004distributed}. Existing distributed source coding methods, such as the lossless Slepian-Wolf coding and the lossy Wyner-Ziv coding, typically require the encoders to have some knowledge of the joint statistics of the sources available at each of the users. Distributed source coding methods enable the users to exploit this correlation between the sources to further compress them without compromising the ability to reconstruct the sources from the compressed representations.

\textbf{Design Considerations}: 
The main distinction between uplink compression for federated learning and conventional source coding  follows from the specific characteristics of the learning setup. In particular, compression methods in federated learning should account for the following considerations:
\begin{enumerate}
    \item The model updates in federated learning are characterized by the lack of a unified statistical model, and the joint distribution of the model updates is not available. 
    \item The model compression procedure is a part of the global aggregation stage, and the compressed updates are to be transmitted through the uplink channel. 
    \item Further to the above, the  encoding  procedure  should  be combined with an appropriate joint decoding procedure, as the server is interested in aggregating the models $\{\boldsymbol{\theta}_t^i\}$, and not in the encoded symbols $\{\boldsymbol{s}_t^i\}$. 
\end{enumerate}

\textbf{Uplink Compression Methods}: 
While the local  encoding stage can be treated as a  source coding setup, distributed source coding techniques are typically not suitable for compressing the model updates. This follows mainly from the lack of a known statistical characterization of the joint distribution of the model updates, which is typically exploited by distributed source coding methods. Furthermore, the fact that in federated learning different users are often randomly selected to  participate in each round (see ``\nameref{ssec:Resource allocation}") and aggregation often involves truncating some of the model updates (see ``\nameref{ssec:truncation}") makes the application of distributed source coding techniques, which relies on the joint decoding of all the coded representations, quite challenging to implement.  Therefore, uplink compression in federated learning is carried out by each user individually in a disjoint manner with the remaining users, and (non-distributed) lossy source coding are typically utilized to achieve highly compressed representation. 

Lossy compression can cause loss of information, and therefore leads to accuracy degradation in the optimization task. 
However, the fact that the updates are compressed for a specific task, i.e., to obtain the global model by, for example, federated averaging, implies that federated learning with bit constraints setup can be treated as a task-based compression scenario. This task is accounted for in the selection of the compression scheme, using one for which the error term vanishes by averaging regardless of the values of $\{\boldsymbol{\theta}_t^i\}$. 

\begin{figure*}
    \centering
    \includegraphics[width=\linewidth]{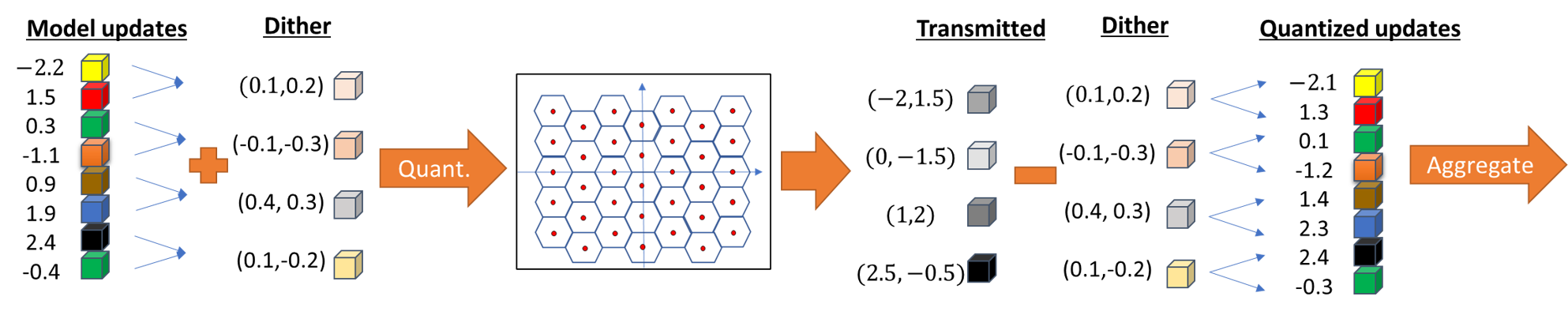}
    \vspace{-0.3cm}
    \caption{Universal quantization via random lattices illustration.}
    \label{fig:SubDithVecQuant}
\end{figure*}

In particular, uplink compression methods in federated learning can be divided  to two general approaches:
\begin{enumerate}
    \item \textit{Sparsification}, where the model updates $\boldsymbol{\theta}_t^i - \boldsymbol{\theta}_{t-E}^i$ are encoded by preserving only a subset of their entries. Sparsification can be implemented based on the value of the model updates, e.g., by nullifying the model updates of lesser magnitude  \cite{stich2018sparsified} or by encoding each layer into a low rank approximation \cite{konevcny2016federatedStr}. The rationale here is that nullifying the weak parameters is expected to have only a minor effect on the accuracy of the  model. Alternatively, one can sparsify the model updates in a random fashion by randomized subsampling and sparifying masks \cite{konevcny2016federatedStr}. The latter strategy relies on the fact that when the masks are randomized in an i.i.d. manner among the users, the global model aggregated via federated averaging will be hardly affected by this sparsification when the number of participating users is sufficiently large. 
    \item \textit{Quantization}, in which each parameter (or set of parameters) are mapped into a finite-bit representation.   
    For example, one-bit quantization can be implemented by merely taking the sign of each model update entry. In fact, when the model updates are the stochastic gradients, namely, each user only implements $\SDGIter=1$ iteration in the local training, one can still guarantee  convergence of the learned model with this strategy, referred to as signSGD \cite{bernstein2018signsgd}. Nonetheless, this crude quantization can notably reduce the convergence rate and the accuracy of the learned model after a finite number of iterations.
    
    Quantization schemes studied in the areas of compression and analog-to-digital-conversion often require knowledge of the distribution of their input to determine which values are mapped to which discrete representation.
    This limits their application in federated learning setups.
    However, by exploiting dithered (randomized) quantization techniques, federated learning of accurate models can be achieved regardless of the distribution of the model updates, e.g., \cite{alistarh2017qsgd}. This follows since the distortion induced by dithered quantization is statistically uncorrelated with the model updates, and thus its effect is mitigated by the averaging operation carried out by the server.  Furthermore, dithered quantization can greatly benefit from having a source of common randomness between the users and server \cite{shlezinger2020uveqfed}, which is a feasible requirement in federated learning setups, as such a shared random number generator allows implementing subtractive dithered quantization  \cite{gray1993dithered}, illustrated in Fig.~\ref{fig:SubDithVecQuant}.
    An example of compression for federated learning in the form of universal dithered quantization is given on Page~\pageref{Box:Uveqfed}.
\end{enumerate} 
Both approaches can be followed by a lossless source code,  exploiting the structures induced by sparification and coarse discretization to further compress the model updates. Sparsified model parameters can also be notably compressed using simple linear projections, from which the sparse vector can be  recovered using compressed sensing mechanisms \cite{amiri2020machine}.

\begin{figure}
    \centering
    \includegraphics[width=\columnwidth]{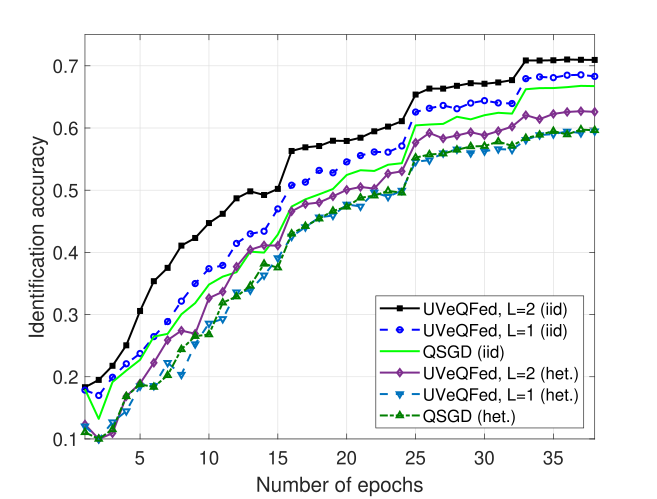}
    \vspace{-0.2cm}
    \caption{Accuracy versus number of epochs of FedAvg combined with universal vector quantization, assigning one bit per model parameter, adopted from \cite{shlezinger2020uveqfed}. The setup considers the training of a 5-layer convolutional neural network using the CIFAR-10 dataset with both i.i.d. and heterogeneous data division among $N=10$ devices. The compared schemes are UVeQFed with 2-dimensional lattices ($L=2$) and 1-dimensional lattices ($L=1$), compared with QSGD \cite{alistarh2017qsgd} which implements the encoding as UVeQFed with $L=1$ without subtracting the dither at the server side.}
    \label{fig:CIFAR_Q2_het}
\end{figure}

\begin{tcolorbox}[float*=t,
    width=2\linewidth,
	toprule = 0mm,
	bottomrule = 0mm,
	leftrule = 0mm,
	rightrule = 0mm,
	arc = 0mm,
	colframe = myblue,
	colback = mypurple,
	fonttitle = \sffamily\bfseries\large,
	title = Dithered Quantization for Federated Learning]	
  Universal quantization refers to discretization methods which are invariant of the statistical model of the discretized quantity, and it is typically desirable to quantize such that the distortion is not determined by the input. A common strategy to quantize in such a universal manner is via probablistic quantization, i.e., via dithering \cite{gray1993dithered}. Applying scalar dithered quantization, namely, adding noise to each element of the model update followed by uniform quantization, results in the QSGD method, proposed in \cite{alistarh2017qsgd}. Applying  subtractive dithered quantization, as proposed in \cite{shlezinger2020uveqfed}, can be achieved for both scalar and vector quantizers by requiring the server to share a source of common randomness, e.g., a random seed, with each user. 
  The universal vector quantization for federated learning (UVeQFed) scheme of  \cite{shlezinger2020uveqfed} consists of the following steps: First, an $L$-dimensional lattice  $\mathcal{L}$ is fixed. Upon local encoding, each device normalizes its local model updates \eqref{eqn:ModelUpdate} by $\ScaleFactor$ times its norm, where $\ScaleFactor>0$ is a given scaling factor. The normalization result is divided into $\Npoints$ $L$-sized vectors, to which the users add  a random dither signal randomized in an i.i.d. fashion from a uniform distribution over the basic cell of the lattice. The dithered signal is discretized by projecting to the nearest lattice point, and the discrete quantity is further compressed prior to transmission using lossless entropy coding. The server decompresses the model updates by recovering the lattice point, and subtracting the dither signal from it, as illustrated in Fig.~\ref{fig:SubDithVecQuant}.  The fact that the users and the server can generate the same dither signals relies upon their ability to share a source of common randomness. When the dither is not subtracted at the server and the lattices are scalar, i.e., $L=1$, UVeQFed specializes QSGD \cite{alistarh2017qsgd}, which is simpler to implement but results in increased distortion.
  
  Using of randomized lattices results in a compression scheme which is simple to implement, invariant of the model updates statistics, and maintains the theoretical performance guarantees of local SGD carried out without any compression. 
      In particular, combining local SGD with UVeQFed achieves the same convergence profile as in \eqref{eq:FedAvg_Convergence} for the strongly-convex smooth objective with bounded gradients scenario (see {\em Convergence of Local SGD} on Page \pageref{Box:FedAvg_convergence}). The effect of the compression mechanism results in $\ConvBound$ replaced with $\ConvBound + 4 \Npoints\ScaleFactor^2 \LatFactor \SDGIter^2  \sum \limits_{i=1}^N p_i^2 \sigma_i^2$, where $\LatFactor$ is the  normalized second order moment of the lattice $\mathcal{L}$. This yields the same asymptotic convergence behavior of $\mathcal{O}(1/T)$ as local SGD without compression. Furthermore, the combination subtractive dithering and multivariate lattice quantization results in improved accuracy of the learned model when training \acp{dnn} with non-convex loss surfaces, as demonstrated in Fig.~\ref{fig:CIFAR_Q2_het}.
\label{Box:Uveqfed}
\end{tcolorbox}	

%
%
\subsection{Privacy Preservation}
\label{ssec:differential_privacy}
Large-scale collection of sensitive data entails risks to the privacy of each individual user. 
Although each local update does not explicitly contain the user’s data, the local updates may be used to reveal the data used for computing them \cite{zhu19deep}.
In general, there are several potential adversaries from which privacy should be preserved in federated learning: the server that may attempt to infer private information using all model updates received; malicious users who can deviate from the protocol execution to achieve  information on data held by honest users (see ``\nameref{ssec:truncation}"); and external adversaries that try to access the private data.

\textbf{Encoding for Privacy}:
A natural strategy to encode the model updates to preserve privacy is to encrypt them. 
Encrypting the model updates implies that they cannot be recovered, and thus cannot be traced back to the data, by external adversaries. In order to preserve privacy while also allowing to combine the encrypted model updates into a global model, one can use secure distributed coding mechanisms, also known as multi-party encryption. In particular, the family of additively homomorphic multi-party encryption schemes allow to perform additive calculations, e.g., averaging-based aggregation, on the encrypted data without having to first decrypt it \cite{fontaine2007survey}. 
The result of the computation is an encrypted form, such that when it is decrypted, the output is the same as if the operations had been performed on the unencrypted data.

Multi-party encryption methods require  the participating entities to share a secret key. An alternative approach to implement secure encoding without key sharing is by randomly obscuring the source via, e.g., artificial noise. This strategy is widely used in physical layer security mechanisms \cite{tsai2014power}, which are well-studied in the signal processing and communication literature. Here, the noise is typically designed to achieve a vanishing mutual information criterion, known as weak secrecy, at the adversary side, i.e., after it has been transmitted. Encoding by artificial noise can also achieve privacy guarantees which are more powerful than the weak secrecy and are invariant of their transmission, such as  differential privacy \cite{dwork2014algorithmic}. The differential privacy level represents how much does the algorithm output change when modifying a single data sample, and thus reflects on the ability to retrace a data sample from the outcome of the training procedure (the exact mathematical formulation of this notion of $(\epsilon,\delta)$ differential privacy is given in the box on Page~\pageref{Box:NbAFL}).
Encoding by artificial noise can guarantee achieving a predefined level of differential privacy, where the noise should be proportional to the sensitivity of the output, i.e., the maximum change of the output due to the inclusion of a single data instance.

\textbf{Design Considerations:} Encoding for privacy allows the users to guarantee that the data used in training their local models cannot be recovered by neither the server nor external adversaries. Nonetheless, federated learning brings forth the following considerations, which should be accounted for in the design of privacy preserving encoding methods:
\begin{enumerate}
    \item Federated learning is not a one-shot operation, but is carried out over multiple rounds, typically with different users participating at each round due to device selection and user dropout. Privacy preserving encoding methods must be thus fully robust to users dropping at any point.
    For instance, the use of secure key distribution protocols requires a pre-defined threshold of users in order to decrypt the global model, implying that dropout may result in inability to decrypt the global model. 
    \item It is desirable not to substantially further increase the communication overload for privacy preservation.
    \item The model updates $\myVec{\theta}_t^i$, which are encoded into  $\myVec{s}^i_t$, are obtained from the local optimization procedure, i.e., the minimization of the local objective using the dataset $\mySet{D}^i$. Thus, the sensitivity to changing a sample may differ depending on the local optimization algorithm.
\end{enumerate}

\textbf{Privacy Preserving Encoding Methods}: 
Existing federated learning approaches to preserve privacy commonly use either secure encryption or artificial noise obscuring.
Each method is characterized by a different trade-off relevant to the unique federated learning  challenges, as discussed in the following:
\begin{enumerate}
    \item {\em Encryption} for federated learning typically relies on additive homomorphic schemes. This approach exploits the fact that the model updates are often aggregated by averaging  to ensure privacy preservation while maintaining the FedAvg flow \cite{aono2017privacy}.
However, the encrypted updates are of a larger size compared to the plain updates, resulting in an increased communication factor.  Additional forms of encryption for federated learning which do not involve homomorphic encryption have been proposed, e.g., \cite{bonawitz2017practical}, though with additional communication overhead and possible limited compliance to the presence of stragglers and limited device participation. In general, encryption methods involve additional communications overhead, as well as possible secret key sharing  between the users \cite{aono2017privacy}.

    \item {\em Artificial noise} implies that each user perturbs its local updates and only sends a randomized version to the server to protect against private information leakage  \cite{wei2020federated}. The common criterion used for characterizing privacy guarantees is differential privacy, being an indicator on the ability to recover a single data sample from the obscured model updates. 
    
    Adding artificial noise for differential privacy considerations naturally gives rise to a trade-off between privacy and convergence performance. 
High noise variance makes it more difficult for attackers to recover the information, but at the same time also degrades the convergence of the learned model.  
Nonetheless, for a given desired privacy level, which dictates the variance of the artificial noise obscuring the model updates, increasing the number of participating users $N$ mitigates the convergence degradation due to privacy preservation, as the effect of the artificial noise vanishes in averaging. Furthermore, it was observed in  \cite{wei2020federated} that for each protection level, there exists a specific finite setting of the number of local iterations $E$ that optimizes the convergence performance for a given protection level.
Thus, the design of privacy-preserving algorithms should be carefully considered to ensure consistency of the optimization problem.

Two popular mechanism for achieving differential privacy in federated learning encode the model updates by corrupting them with Laplacian and Gaussian artificial noise signals \cite{wei2020federated}. 
    Differential privacy can also be achieved with non-artificial noise through the channel noise induced in uplink transmission when using uncoded transmissions, as  shown in \cite{liu2020privacy}.  
An example of a differential privacy scheme is detailed under {\em Privacy Boosting by Noising Before Model Aggregation}  on Page~\pageref{Box:NbAFL}.
\end{enumerate}
Finally, it is noted that artificial noise obscuring and encryption can also be combined in  a hybrid manner to jointly guarantee privacy preservation in federated  
learning, as proposed in \cite{truex2019hybrid}.

\begin{figure*}[!t]
\centering
\subfigure[Privacy boosted federated learning via NbAFL.]
{ 
\label{fig:NbAFLflow} 
\includegraphics[height=5.8cm, width=0.35\linewidth]{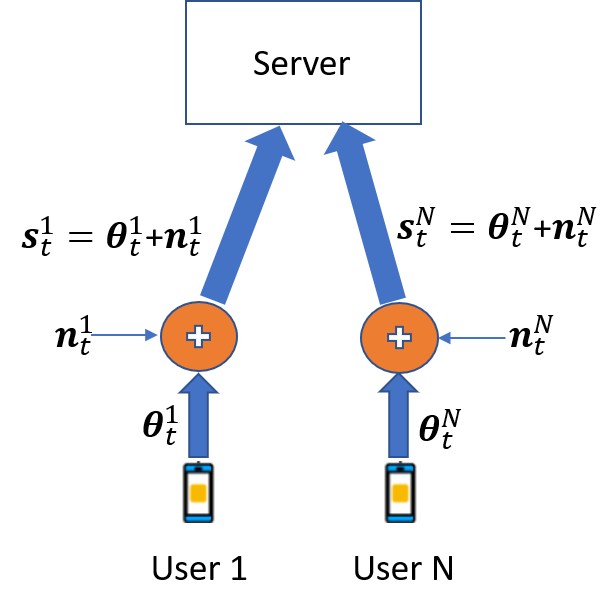}}
\quad
\subfigure[Convergence profile of NbAFL.]{ 
\label{fig:NbAFLsim} 
\includegraphics[height=6cm, width=0.4\linewidth]{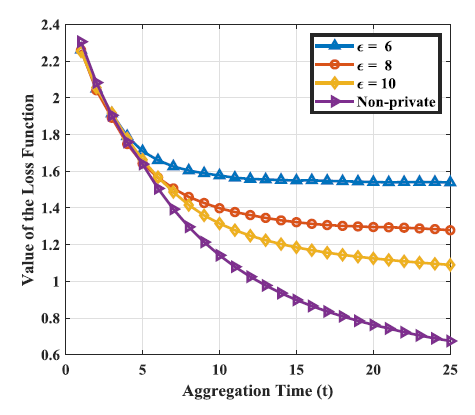}} 
 \caption{\label{fig:NbAFLscheme} NbAFL scheme proposed in \cite{wei2020federated} for federated learning based differential privacy with Gaussian artificial noise. The numerical study in (b), adopted from \cite{wei2020federated} with the authors' permission, considers the federated learning on the standard MNIST dataset for handwritten digit, where the loss function is compared under different protection levels $\epsilon$.}
 \vspace{-0.3cm}
\end{figure*}

\begin{tcolorbox}[float*=t,
    width=2\linewidth,
	toprule = 0mm,
	bottomrule = 0mm,
	leftrule = 0mm,
	rightrule = 0mm,
	arc = 0mm,
	colframe = myblue,
	colback = mypurple,
	fonttitle = \sffamily\bfseries\large,
	title = Privacy Boosting by Noising Before Model Aggregation]	

    Consider a  federated learning system consisting of one server and $N$ users, where each user holds a dataset with size $|\mySet{D}^i| = n_i$. The server and users are assumed to be honest, however there are external adversaries targeting at users' private information.
    There are at most $\ell (\ell \leq T)$ exposures of uploaded parameters from each user in the uplink where $T$ is the number of aggregation times.
    The goal is to implement federated learning system subject to an $(\epsilon, \delta)$- DP constraint defined as:

    \textit{A randomized mechanism $\mathcal{M}:\mathcal{X} \rightarrow \mathbb{R}$ satisfies $(\epsilon,\delta)$-DP, if for all measurable sets $\mathcal{S} \subseteq \mathbb{R}$ and for any two adjacent datasets $\mySet{D}^i,\tilde{\mySet{D}}^i \in \mathcal{X}$}: 
        $\mathbb{P} (\mathcal{M}(\mySet{D}^i) \in \mathcal{S}) \leq e^\epsilon \mathbb{P} (\mathcal{M}(\tilde{\mySet{D}}^i) \in \mathcal{S}) + \delta $. 
        
The  Noising before Model Aggregation Federated Learning (NbAFL) scheme proposed in \cite{wei2020federated} uses the Gaussian mechanism to guarantee $(\epsilon, \delta)$-DP, i.e., the mapping (\ref{eq:precoding}) is in the form of 
\begin{equation}
    \boldsymbol{s}_t^i = \boldsymbol{\theta}_t^i + \boldsymbol{n}_t^i,
\end{equation}
where $\boldsymbol{n}_t^i$ is an independent Gaussian noise, whose elements are i.i.d drawn from $\mathcal{N}(0,\sigma_i^2)$.
To achieve the differential privacy constraints using the Gaussian mechanism, it is required to choose  $\sigma_i \geq c \Delta s^i / \epsilon$ and the constant $c \geq \sqrt{2 \ln (1.25 / \delta)}$ where $\epsilon \in (0,1)$ \cite{dwork2014algorithmic}, and $\Delta s^i = \max_{\mySet{D}^i,\tilde{\mySet{D}}^i} ||s(\mySet{D}^i) - s(\tilde{\mySet{D}}^i)||$ is the local sensitivity. 
Combining this with the result from \cite{dwork2014algorithmic} given above, the sufficient variance noise for the users, $\sigma_i$, 
is obtained by:
\begin{equation}
\label{eq:variance_noise}
  \sigma_i = \frac{4 c \ell C}{\epsilon \cdot \min \{n_i\}}, \forall{i}  
\end{equation}

As shown  in \cite{wei2020federated}, the expected gap between the achieved loss function by NbAFL and the minimum loss is a decreasing function of $\epsilon$. This illustrates the trade-off between privacy and accuracy: by increasing $\epsilon$, the degree of privacy is decreased, however, this results with lower noise variance as can be seen from equation (\ref{eq:variance_noise}), hence the performance of the algorithm improves, as illustrated in Fig. \ref{fig:NbAFLsim}.

\label{Box:NbAFL}
\end{tcolorbox}

%
%
%
%

\section{Uplink Transmission}
\label{sec:Transmission}
The second stage in the global aggregation step of federated learning involves the transmission of the encoded updated models, produced as the outputs of the local encoding stage, to the server. This operation is essentially a communications procedure, as it requires multiple devices to convey information to a remote server in a reliable manner and with minimal delay.
%

%
%
Here, we  
present the conventional transmission strategy, and discuss the challenges which arise from this approach. We then discuss signal processing and communication methods for dealing with similar challenges, and elaborate on their adaptation for the federated learning setup. 

Federated learning typically involves a large number of edge users, which often reside in various physical locations, and utilize different infrastructures for communicating with the server. In particular, some of the users may communicate over the same shared  channel, e.g., a set of mobile devices served by the same wireless access point, while some can utilize non-shared wireline links, as illustrated in Fig.~\ref{fig:Constraints}. 
As we discuss in the sequel, the main challenges associated with the uplink transmission stage arise when the users communicate over a shared medium, e.g., when the users are wireless transceivers served by the same access point. Therefore, to highlight the potential benefits of properly designed  uplink transmission schemes, we focus henceforth on the extreme scenario in which all the users communicate with the server over the same wireless channel. Nonetheless, the methods detailed next can be adapted and contribute to federated learning networks where distinct subsets of the users share the uplink channel.

\textbf{Conventional Transmission}: 
From an implementation perspective, it is often simpler to separate the communication functionality and the learning application. In this case, the learning procedure is agnostic of how the model updates are communicated, while relying on an existing communication architecture to provide these transmissions. 
Conventional communication architectures are ignorant of the learning task, and  enable multiple users to share the channel by mitigating interference via resource allocation. This is achieved by dividing the channel resources into multiple resource blocks and assigning each block to a different user. A common approach to do so is by splitting the available bandwidth into non-overlapping frequency bins via orthogonal frequency division multiplexing multiple access protocols.  Such orthogonlization results in each user having a separate channel with the server, i.e.,   $\mathbb{P}(\myVec{y}_t^1,\ldots,\myVec{y}_t^N|\{\myVec{x}_t^i\}) = \prod\mathbb{P}(\myVec{y}_t^i|\myVec{x}_t^i)$. Then, reliable communications is achieved by having each user utilize a channel code with code-rate not larger than the capacity of its individual channel, guaranteeing that the server can accurately recover $\myVec{s}_t^i$ from $\myVec{y}_t^i$ with arbitrarily high probability.

While the conventional strategy allows reliable transmission of the model updates from the users to the server, it can induce a notable communication delay. This delay is sometimes more dominant than the duration it takes for each user to carry out its local optimization, thus  affecting the convergence of the learned global model. To see this, recall that wireless channels are limited in bandwidth and energy, and thus the achievable rate of each user decays with the number of participating users, as the channel resources are divided among more users. Since the number of devices taking part in the federated learning procedure is typically  very large, while the model updates are often comprised of a massive amount of parameters,  the delay of each user can be quite significant. Furthermore, the delays may vary considerably between different users due to the heterogeneous nature of edge devices, notably affecting the overall delay of each uplink transmission round, being dictated by the user with the largest delay. As federated learning is comprised of multiple rounds, the conventional strategy is likely to result in lengthy convergence time of the learning procedure, which is translated into a degraded accuracy of models being learned in a given finite time.

{\textbf{Signal Processing Perspective}}: Mapping the encoded model updates $\myVec{s}_t^i$ into a channel input $\myVec{x}_t^i$ is comprised of two main tasks: the first is the traditional communication operation of formulating a channel input which achieves reliable communication at the maximal rate given the assigned channel resources via, e.g., modulation and channel coding; the second task, which encapsulates much of the potential of minimizing the transmission delay, is problem of {\em resource allocation} \cite{meshkati2007energy}. Resource allocation is well studied in the signal processing and communication literature, considering various objectives including  throughput and energy efficiency.  However, unlike conventional communication systems where the receiver is interested in recovering the transmitted messages, in federated learning the server needs the model updates only as an intermediate step in producing the global model. This gives rise to learning-aware resource allocation and transmission schemes, detailed in the following. In particular, we first discuss how the resources of the wireless channels can be divided among the users while accounting for the learning task. Then  we present the notion of over-the-air federated learning, allowing the users to share the same resources, i.e., communicate in a non-orthogonal fashion, while treating the effect of the channel as a form of functional network over-the-air computation \cite{liu2020over}.

%
%
\subsection{Learning-Aware Resource Allocation}
\label{ssec:Resource allocation}
The uplink channel is comprised of multiple resource blocks. These blocks represent, e.g., portions of the bandwidth and temporal transmission slots. While dividing these resources equally among all users is likely to limit the convergence rate of the learned global model, one can notably improve the convergence rate by dividing the resources in a manner which accounts for the overall learning procedure. 

\textbf{Resource Allocation in Wireless Networks}: 
Allocating the resources in wireless networks with multiple users is comprised of {\em user selection} and {\em resource management}: The former deals with the selection of the users to whom resources are allocated, i.e., the users which are allowed to transmit; The latter refers to the division of the resources among the selected users. In the conventional signal processing literature on resource allocation, which deals with communications over wireless networks not necessarily for the purpose of training a parametric model, user selection and resource management are commonly studied jointly. Common resource allocation strategies are categorized into static pre-defined allocation and dynamic divisions\myCite{zhao2007survey}. Leading design tools include optimization-based strategies, which formulate the resource allocation problem using a network utility function,  and game-theoretic approaches that utilize a local utility for each user \cite{meshkati2007energy}. These utility measures typically characterize communication rate, fairness, and energy efficiency. Here, partial  participation, i.e., user selection, is obtained as a byproduct of this utility, which may result in some users not utilizing any  channel resources.

\textbf{Design Considerations}:
As opposed to resource allocation in conventional wireless networks, partial user participation is typically desirable in federated learning  regardless of the availability of channel resources. As a result, user selection and resource management are often considered as separate tasks, as opposed to the conventional resource allocation literature.

{\em User Selection}: the need to limit the amount of users that participate in each global aggregation round was already identified in the original federated learning paper by McMahan et al. \cite{mcmahan2017communication}. The motivation for user selection stems not only from the communication bottleneck, but also from reasons associated with computation and load balancing. It was noted that increasing the number of participating users beyond some point no longer contributes to the learned model \cite{mcmahan2017communication}, and that an accuracy similar to that achieved using full device participation can be obtained while having each user participate in a subset of the global aggregation rounds \cite{li2019convergence}. These observations capture the main distinction between the communications problem of uplink transmission, in which the goal is to convey the messages from all the users to the server, to uplink transmission in federated learning, where messages are useful only if they contribute to the learned global model.

The user selection problem deals with determining the set of participating users, denoted $\mathcal{G}_t$. Each user which is not part of this set does not participate in the aggregation at iteration $t$, and thus:
\begin{equation}
\label{eq:UserSel1}
    i \notin \mathcal{G}_t \Leftrightarrow \varphi_i(\cdot) \equiv 0.
\end{equation}
The design of a user selection method must account for the following considerations:
\begin{enumerate}
    \item Knowledge of  $\mathcal{G}_t$ is essential for computing the averaged global model via \eqref{eq:FedAvg_Server}. Consequently, $\mathcal{G}_t$ is typically determined at the server side, and is distributed in the downlink transmission, implying that users which are not selected do not need to carry out  local training.
    \item As the essence of user selection is to limit the amount of participating users, it must result in the cardinality of $\mathcal{G}_t$ being bounded, i.e., $|\mathcal{G}_t|<K$ for each $t \in \mySet{I}_T$ and for some $K\leq N$, preferably $K \ll N$. 
    \item In order to allow all users to participate in the learning procedure, user selection should also involve a scheduling policy\myCite{yang2019scheduling}, such that $\cup_{t \in\mySet{I}_T}\mathcal{G}_t = \mathcal{N}$, where the equality either holds directly or with high probability.
\end{enumerate}

{\em Resource Management}:
Once the set of participating users is determined, the channel resources are allocated among these users. As opposed to user selection, which is often utilized regardless of whether the users share the same channel or not, resource management is relevant only for users which communicate over the same media.  In particular, resource management deals with the division of the channel resource blocks, e.g., frequency bins, among the users in $\mathcal{G}_t$. The resource blocks allocated at iteration $t$ to user $i \in \mathcal{G}_t$ determine its transmission mapping $\varphi_i(\cdot)$ via, e.g., its spectral support.

Schemes for managing the channel resources should account for the following considerations:
\begin{enumerate}
    \item Resource reuse, i.e., assigning the same resource block to more than one user, inevitably yields some level of interference. While cross interference can sometimes be exploited, as detailed in the ``\nameref{ssec:Over-the-air FL}" subsection, it is often preferred to avoid such reusing, and thus the number of resource blocks should not be smaller than $|\mathcal{G}_t|$.
    \item The participating users in federated learning are often heterogeneous in their system capabilities and energy, motivating a non-even division of the channel resources.
    \item The objective of resource management is to allow rapid and reliable transmission of the  messages $\{\myVec{s}_t^i\}_{i \in \mathcal{G}_t}$. Consequently, the measure of interest here is the maximal delay in conveying these fixed-length messages, which is fundamentally different from typical communications measures such as achievable sum-rate. 
\end{enumerate}

\textbf{Resource Allocation Methods}: 
Based on the above considerations, one can mitigate the effect of some of the challenges associated with federated learning  by properly selecting the users which participate in each round, and then divide the channel resources among them in a learning-aware manner. While the objectives may differ from those used for conventional resource allocation, one can still utilize existing methods with proper adaption to account for the learning task.
Resource allocation methods for federated learning include:
\begin{enumerate}
    \item {\em User selection} can be either carried out via deterministic scheduling, random selection, or using an online policy. For instance, a deterministic scheduling policy can assign the most computationally weak devices to participate in the same round, thereby reducing the delay induced by device heterogeneity by decreasing the local processing delay of the remaining rounds. 
    
    In random selection, a fixed amount of $\NSelUsers \leq N$ users are randomized, such that $|\mathcal{G}_t| = \NSelUsers$ and $\{\mathcal{G}_t\}_{t\in \mathcal{I}_T}$ are mutually independent and identically distributed. 
    A simple strategy is to implement such random selection in a uniform fashion, i.e., having $\mathcal{G}_t$ randomized uniformly from all subsets of $\mathcal{N}$ of cardinality $K$ \cite{mcmahan2017communication,bonawitz2019towards}. This simple strategy, which treats all users in a similar manner, can be proven to yield a similar asymptotic convergence profile to that of FedAvg with full device participation \cite{li2019convergence}. 
    Nonetheless, one can incorporate some knowledge about the characteristics of each user and its contribution to the training procedure by deviating from uniform selection. For instance, by modifying the distribution based on which $\mathcal{G}_t$ is randomized from $\mathcal{N}$ to account for the contribution of each user to the learning procedure and its expected delay, one can reduce the convergence delay without affecting accuracy compared to full device participation \cite{chen2021communication,chen2019joint}. For a detailed example, see {\em Delay Minimization with Probabilistic User Selection} on Page~\pageref{Box:Delay}. 
    An alternative non-uniform approach is to use importance sampling to adaptively assign  probabilities to users in each  round. It was shown in \cite{balakrishnan2021resource} that implementing random user selection via importance sampling  based on the users’ computing, communication, and data resources, can notably increase the convergence speed.
    
    Online scheduling selects the participating users in each round by accounting for the past selected users and their observed behavior, e.g., delay. Such selection policies can be designed by adopting a multi-armed bandit framework \cite{xia2020multi}. Here, the users  are selected to strike a balance between the exploitation of users which performed well in the past, e.g., transmitted with low delay, and the exploration of users which have not sufficiently participated until the current round. 
    For instance, the work \cite{xia2020multi} assigned each user with an index, which is comprised of the average transmission time of the user based on previous rounds (exploitation term), and a second term which is inversely proportional to the number of rounds the user transmitted (exploration term). The latter ensures that users who have transmitted less are given priority, in order to capture potential environments changes that may reduce their transmission delay. In each round, the server selects the set of $K \leq N$  users with the highest indexes.

    \item {\em Resource management} can be implemented in a centrally-controlled manner, optimizing a centralized utility as in conventional resource allocation. This utility function accounts for measures of the learning task, e.g., model convergence and training time \cite{dinh2019federated}. Here, the allocation of the resource blocks can be either dictated by the server or by the wireless access point via which the sharing users communicate with the server, and conveyed to the users in the model distributing step.  An example of a resource management scheme which accounts for delay minimization is detailed under {\em Delay Minimization with Probabilistic User Selection} on Page \pageref{Box:Delay}. Alternatively, one may allow the users to utilize the channel resources in a decentralized opportunistic manner, using, e.g., dynamic spectrum access techniques \cite{song2021federated}. 
    
\end{enumerate}

\begin{figure*}[!t]
\centering
\subfigure[Accuracy versus number of iterations]{ 
\label{fig1b} 
\includegraphics[width=0.42\linewidth]{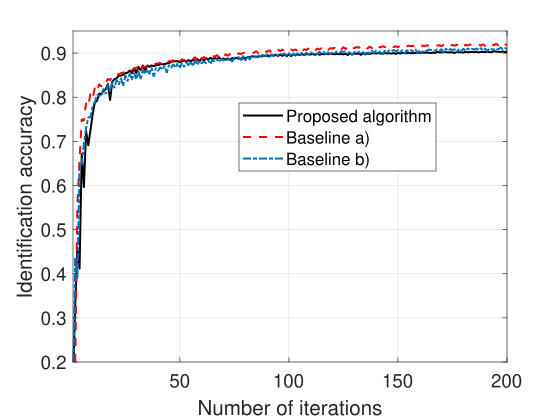}}
\quad
\subfigure[Convergence time versus number of selected devices]{ 
\label{fig1c} 
\includegraphics[width=0.42\linewidth]{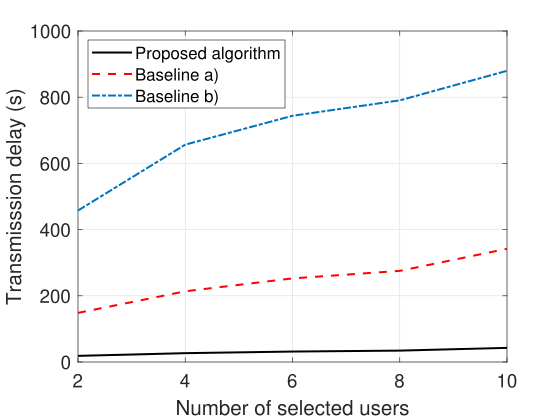}} 
 \caption{\label{fig:DelaySim} Federated learning for hand-written digit recognition from the MNIST dataset distributed among $N=15$ devices, adopted from \cite{chen2021communication}. 
 Here, the model trained is comprised of fully connected neural network with a single hidden layer of $50$ neurons. Here, both {\em Proposed algorithm} and {\em Baseline A)} refer to resource allocation based on delay minimization (see box on Page \pageref{Box:Delay}), while the former is combined with model compression based on \cite{shlezinger2020uveqfed}; {\em Baseline b)} refers to random resource allocation.}
 \vspace{-0.3cm}
\end{figure*}

\begin{tcolorbox}[float*=t,
    width=2\linewidth,
	toprule = 0mm,
	bottomrule = 0mm,
	leftrule = 0mm,
	rightrule = 0mm,
	arc = 0mm,
	colframe = myblue,
	colback = mypurple,
	fonttitle = \sffamily\bfseries\large,
	title = Delay Minimization with Probabilistic User Selection]	
    Consider a set of users which communicate with the server via an access point over a shared wireless channel comprised of $K$ distinct resource blocks of bandwidth $\Bandwidth$.  When the $k$th resource block is allocated solely to user $i \in \mathcal{N}$, it communicates with the server over a  flat fading channel with Gaussian noise and interference, given by
    \begin{equation}
        \label{eqn:Channel}
        y_i = h_i x_i + w_i + v_k.
    \end{equation}
    Here, $x_i$ is the channel input with maximal transmit power $P$; $h_i$ is the channel gain of the $i^{\text{th}}$ device, satisfying  $h_i\propto d_i^{-2}$ where $d_i$ is its distance to the access point; $w_i$ is white Gaussian noise with variance $\sigma_w^2\Bandwidth$; and $v_k$ is the interference in the $k$th resource block, whose energy is $\sigma_{v,k}^2$. The data rate over the channel \eqref{eqn:Channel} is given by $R_{i,k}=\Bandwidth \log_2 \big(1 + \frac{h_i P}{\sigma_{v,k}^2 +\sigma_w^2\Bandwidth}  \big)$. 
    
    The learning-aware scheme proposed in \cite{chen2021communication} allocates the resources for such uplink transmissions in  two stages: 
    
    {\em Probabilistic user selection:} First, $K$ users are randomly selected to participate in each round. The randomization accounts for both the contribution of each user via the norm of its model updates, as well as its transmission delay, making devices with stronger links more likely to participate. Letting $\alpha_t \in [0,1]$ be a parameter balancing these contributions, the participation probability of user $i$ at round $t$ is computed as
    \begin{equation}
        \label{eqn:Probability1}
        \rho_{t}^i = \alpha_t \frac{\|\boldsymbol{\theta}_{t}^i - \boldsymbol{\theta}_{t-\SDGIter}^i \|}{\sum_{i=1}^{N}\|\boldsymbol{\theta}_{t}^i - \boldsymbol{\theta}_{t-\SDGIter}^i \|} 
        + (1-\alpha_t) \frac{\max_j d_j - d_i}{N\max_j d_j - \sum_{i=1}^{N}d_i}.
    \end{equation}
    The probability vector $[ \rho_t^1,\ldots  \rho_t^N]$ is used to randomize the $K$ participating users $\mathcal{G}_t$.
    
     {\em Delay minimizing resource division}  Next, the resource blocks are divided among the users in  $\mathcal{G}_t$ by minimizing the transmission delay. By letting $\Bits_{t}^i$ be the bits used for representing the encoded model update $\myVec{s}_t^i$, the delay of user $i \in \mathcal{G}_t$ when assigned the $k$th block is given by $\Bits_{t}^i / R_{i,k}$. Consequently, the  resource blocks are allocated in order to minimize the maximal delay via the following constrained optimization problem: 
    \begin{equation}
        \label{eqn:ConstProblem}
        \min_{\{\chi_{t,k}^i\}_{k=1}^K \in \{0,1\}^K} \max_{i \in \mathcal{G}_t} \frac{\Bits_{t}^i}{\sum_{k=1}^{K} \chi_{t,k}^i R_{i,k}}; \qquad {\text{subject to }} \sum_{i \in \mathcal{G}_t} \chi_{t,k}^i = \sum_{k=1}^{K} \chi_{t,k}^i = 1.
    \end{equation}
    The constraint in \eqref{eqn:ConstProblem} guarantees that each resource block is assigned to a single distinct user. As shown in \cite{chen2021communication}, the optimization in \eqref{eqn:ConstProblem} can be recast as an integer linear programming problem, facilitating its solution numerically. This combination of contribution-and-delay-aware user selection along with delay-minimizing resource allocation enables the learning of accurate models at notably reduced delay compared to uniform random user selection, as illustrated in Fig.~\ref{fig:DelaySim}.
\label{Box:Delay}
\end{tcolorbox}

%
%
\subsection{Over-The-Air Federated Learning}
\label{ssec:Over-the-air FL}
An alternative strategy for uplink transmissions in federated learning over shared channels, which can be treated as an extreme case of resource management, is to allow full reuse of all the channel resource blocks. Here, the users simultaneously utilize the complete temporal and spectral resources of the uplink channel in a non-orthogonal manner. The rationale  is to exploit the inherent aggregation carried out by the shared channel as a form of over-the-air computation \cite{liu2020over}. 

\textbf{Over-The-Air Functional Computation}:
The concept of over-the-air functional computation originates in coding studies for multiple access channels  \cite{nazer2007computation}.
It is based on the counter-intuitive finding that interference can be harnessed to help computing, when there is a one-to-one map between a desired computation and a linear function of the transmissions. 
Since the emergence of over-the-air functional computation in multi-user communications, it has been traditionally studied and developed for statistical inference in wireless sensor networks.

\begin{figure*}[!t]
\centering
\subfigure[Over-the-air federated learning with time-varying precoding marked in green fonts.]{ 
\label{fig:COTAFSim1} 
\includegraphics[width=0.6\linewidth]{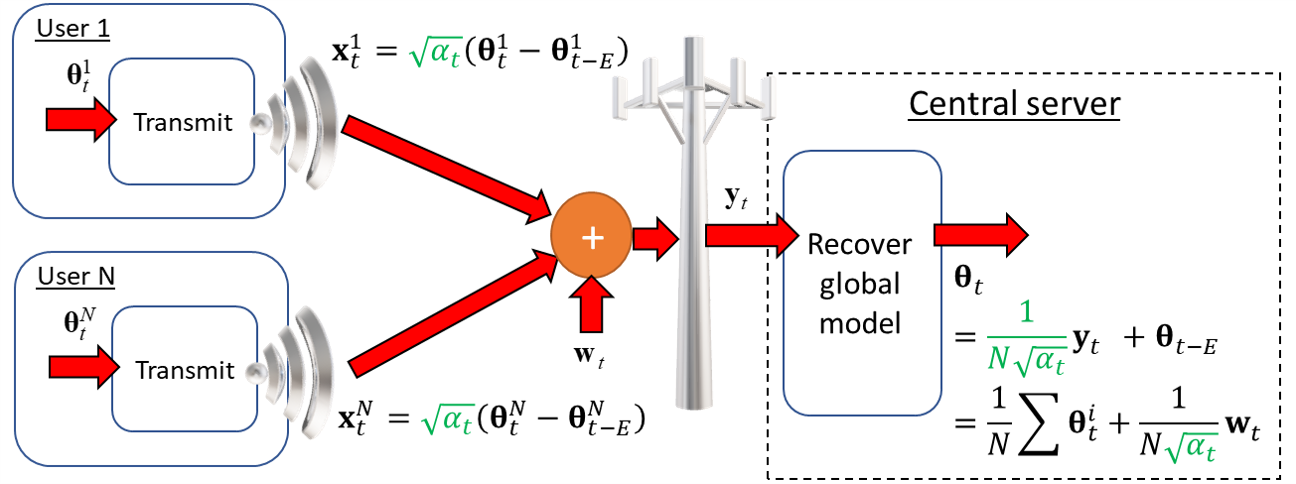}}
\quad
\subfigure[Convergence profile of COTAF.]{ 
\label{fig:COTAFSimb} 
\includegraphics[width=0.3\linewidth]{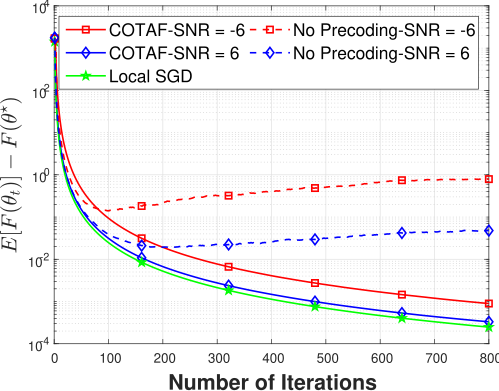}} 
 \caption{\label{fig:COTAFSim} Over-the-air federated learning illustration. The numerical study in (b) considers the federated learning of a linear predictor using the Million Song dataset, where the COTAF scheme of \cite{sery2021over} is compared to noise-free local SGD over orthogonal channels, as well as over-the-air transmission with no time-varying precoding.}
 \vspace{-0.3cm}
\end{figure*}

Over-the-air computation relies on having the users share the same channel, as illustrated in Fig.~\ref{fig:COTAFSim1}. The main advantage of this approach is  that it allows each transmitter to utilize the complete available channel resources, e.g., bandwidth, regardless of the number of users \cite{liu2020over}. For the interference result in some desired linear function over a continuous space, such as the averaging of real-valued vectors, the transmitters should utilize analog signalling. In such cases, the channel input is a continuous function of the model updates, e.g., 
\begin{equation}
\label{eqn:Scaling}
 \myVec{x}_t^i = \varphi_i(\myVec{s}_t^i) = \sqrt{\alpha_t^i}    \myVec{s}_t^i,
\end{equation}
for some power scaling parameter $\alpha_t^i>0$. Using \eqref{eqn:Scaling}, the channel output represents some form of weighted averaging of the transmitted $\{\myVec{s}_t^i\}$. Consequently, transmission is carried out  without converting $\{\myVec{s}_t^i\}$, e.g., the model updates, into discrete coded symbols which should be decoded at the receiver side, as is commonly the case in digital communications. 

\textbf{Design Considerations}:
Over-the-air computations can notably increase the throughput and reduce the transmission delay in federated learning over wireless channels compared with orthogonal division of the channel resources. However, analog signalling induces challenges on the communication methods; For instance, it requires accurate synchronization among all users. Furthermore, in order for the users and the server to guarantee that the combining carried out by the shared channel can be transformed into a desired aggregation mapping,  knowledge of the channel input-output relationship is typically required. 
Consequently, the design of over-the-air computation methods for federated learning must account for the following considerations:
\begin{enumerate}
    \item The desired global combining mapping should be one which can be obtained via over-the-air computations. In particular, for linear uplink channel models where the contribution of the interference is additive, this implies that the desired aggregation mapping should be a nomographic function. 
    Such multivariate functions can be represented as a univariate functional of a combination of their variables, as is the case for FedAvg aggregation \eqref{eq:FedAvg_Server}. This limits the combination of over-the-air computation with non-nomographic aggregation rules, such as trimmed mean (see ``\nameref{ssec:truncation}"). 
    \item The transmission mapping $\varphi_i(\cdot)$ and the post-processing carried out at the server side should mitigate the effect of the channel noise on the learning algorithm.  This is necessary to allow training procedure to converge despite the presence of noise in the model updates.  
    \item Dedicated mechanisms should be introduced to achieve synchronized transmissions.  
\end{enumerate}
The second consideration should be carefully examined for each model, since the sensitivity of  optimization algorithms is highly dependent on their objective. For instance, \ac{sgd}-based optimization over convex objectives is highly affected by noisy observations, where convergence can only be guaranteed to some environment of the optimal solution. However, for non-convex objectives, \ac{sgd} benefits from a minor level of noise, which contributes to its convergence by providing means of avoiding local minimas \cite{Guozhong1995NoiseBackprop}. 
To deal with the third requirement, one can incorporate mechanisms utilized in digital communication protocols, which typically use beacon transmissions and slotted superframes to achieve synchronization. An alternative approach to achieve synchronization is to utilize dedicated schemes designed for over-the-air computations. These include the analog modulation method proposed in \cite{goldenbaum2013robust}, or the broadcasting of a shared cloak during the model distributing step as proposed in \cite{abari2015airshare}.

\textbf{Over-The-Air Federated Learning Methods}: 
The fact that  the model updates in federated learning are typically transmitted in order to be averaged makes over-the-air computation an attractive technique for increasing the throughput when operating over the same shared channel. 
The straight-forward application of over-the-air federated learning in which the transmitters send their stochastic gradients (i.e., local \ac{sgd} with $\SDGIter=1$) precoded to meet a given power constraint was shown to result in low convergence delay when learning is carried out over identical unfaded channels with homogeneous datasets  \cite{amiri2020machine}.  Nonetheless, in more involved setups of federated learning over wireless channels, which include fading, heterogeneous data, and multiple local \ac{sgd} steps, one can notably improve the accuracy of the learned model and the convergence delay by utilizing dedicated precoding mechanisms, as well as integrating methods from the areas of compressed sensing and multi-user \ac{mimo} communication:
\begin{enumerate}
    \item {\em Dedicated precoding} can guarantee convergence of over-the-air local \ac{sgd}  carried out over heterogeneous datasets and with multiple local iterations, i.e., $\SDGIter > 1$. This is achieved by setting the precoder to  account for the fact that the  difference in each set of $\SDGIter$ local iterations is expected to gradually decrease over time \cite{sery2021over}. Building upon this insight, the model updates are scaled by their maximal expected norm, along with a corresponding aggregation mapping at the server side, which jointly result in an equivalent model where the effect of the noise induced by the channel is mitigated over time. An example of an over-the-air federated learning method which enables high throughput signaling with guaranteed convergence by gradual mitigation of the effect of the channel noise  is detailed under {\em Time-Varying Precoding for Over-the-Air Federated Learning} on Page~\pageref{Box:COTAF}.
    
    Additionally, one can precode to mitigate the effect of channel fading of the learned model. For instance, when channel state information  is available at the users, they can perform channel inversion precoding, i.e., to scale the model updates by the channel coefficient \cite{zhu2019broadband}. 
    In practice,  some channels are likely to exhibit deep fading, rendering channel inversion inefficient under a power constraint, and a truncated version of channel inversion is preferable \cite{sery2021over}. Here, only users whose channels attenuation is below predefined threshold transmit, resulting in an additional form of user selection based on channel quality.
    
    \item {\em Compressed sensing} tools allow exploiting the fact that the model updates typically approach being sparse as the training algorithm progresses \cite{amiri2020machine}. Consequently, when the model updates have a similar sparsitiy pattern, e.g., when training using homogeneous datasets, the recovery of the aggregated model can be viewed as the reconstruction of a sparse vector. This may be exploited by applying random compressive linear precoding, which can be applied in fading channels without requiring the users to have channel state information, having the receiver recover the aggregated model via compressed sensing mechanisms \cite{amiri2020federated}. 
    
    \item {\em Multi-user \ac{mimo}} methods, which are applicable when the receiver has multiple antennas, can further facilitate high throughput over-the-air federated learning by exploiting the spatial diversity of such channels. This can be realized by introducing learning-aware beamforming techniques \cite{yang2020federated}, possibly combined with user selection, guaranteeing that in each round users sharing a similar relative angle to the receiver take part in global aggregation. Furthermore, when the receiver has a large number of antennas, one can mitigate the need for accurate channel knowledge in over-the-air federated learning by utilizing massive \ac{mimo} schemes \cite{amiri2020blind}.
\end{enumerate}

\begin{tcolorbox}[float*=t,
    width=2\linewidth,
	toprule = 0mm,
	bottomrule = 0mm,
	leftrule = 0mm,
	rightrule = 0mm,
	arc = 0mm,
	colframe = myblue,
	colback = mypurple,
	fonttitle = \sffamily\bfseries\large,
	title = Time-Varying Precoding Over-the-Air Federated Learning]	
    Consider an uplink wireless channel shared by $N$ users. When the users share the complete channel resources, i.e., communicate in a non-orthogonal fashion, the channel output at round $t$ is related to the channel inputs $\{ \myVec{x}_t^i\}$ via
    \begin{equation}
        \label{eqn:MAC1}
        \myVec{y}_t = \sum_{i=1}^{N}\myVec{x}_t^i + \myVec{w}_t.
    \end{equation}
    Here, the $\myVec{w}_t$ is an additive Gaussian noise vector whose entries have  zero-mean and variance $\sigma_w^2$; $\myVec{x}_t^i$ is subject to an energy constraint $P$; and all vectors are comprised of $d$ elements, i.e., same as the number of model parameters. 

    The inherent averaging carried out by the shared channel \eqref{eqn:MAC1} can be exploited to enable over-the-air federated learning with local \ac{sgd} optimization, as proposed in, e.g., \cite{amiri2020machine,sery2021over}. In this case, the channel input is set to a precoded version of the model updates, given by $\myVec{x}_t^i  = \sqrt{\alpha_t}(\myVec{\theta}_t^i - \myVec{\theta}_{t-\SDGIter}^i)$, where $\alpha_t$ is a precoding parameter, which can change between iterations. Under such a setting, the server computes the updated global model via 
    \begin{equation}
    \label{eqn:GlobalModel}
        \myVec{\theta}_t = \frac{1}{\sqrt{\alpha}_t}\myVec{y}_t + \myVec{\theta}_{t-\SDGIter} 
                        = \frac{1}{N} \sum_{i=1}^{N}\myVec{\theta}_t^i + \frac{1}{N\sqrt{\alpha}_t}\myVec{w}_t.
    \end{equation} 
    This  transmission scheme is illustrated in Fig.~\ref{fig:COTAFSim1}.
    The work \cite{sery2021over} proposed to set the precoder to $\alpha_t = P / \max_i \mathds{E}\big\{\|\myVec{\theta}_t^i - \myVec{\theta}_{t-\SDGIter}^i \|^2\big\}$.
    This precoding term accounts for the fact that the magnitude of the model updates is expected to decrease as the training process progresses, and thus the setting of $\alpha_t$ results in the global model $ \myVec{\theta}_t$ in \eqref{eqn:GlobalModel} being given by $\frac{1}{N}\sum_{i=1}^{N}\myVec{\theta}_t^i $, i.e., the desired federated averaging term \eqref{eq:FedAvg_Server} for $p_i = \frac{1}{N}$ and $\mathcal{G}_t = \mathcal{N}$, corrupted by an additive noise term whose contribution gradually decays with the iteration $t$. 
    
    The time-varying precoding scheme of \cite{sery2021over} is referred to as convergent over-the-air federated learning (COTAF). This is because for the strongly-convex smooth objectives with bounded gradients (see {\em Convergence of Local SGD} on Page \pageref{Box:FedAvg_convergence}), COTAF results in the same convergence profile as in \eqref{eq:FedAvg_Convergence}, with $\ConvBound$ replaced with $\ConvBound + \frac{4d\SDGIter^2G^2\sigma_w^2}{PN^2}$. This implies that the same asymptotic convergence behavior of $\mathcal{O}(1/T)$ is achieved as in orthogonal transmissions over noise-free channels as illustrated in Fig.~\ref{fig:COTAFSimb}, while mitigating the harmful effect of the channel noise and communicating at higher throughput due the full reuse of the channel resources. 
\label{Box:COTAF}
\end{tcolorbox}

%

\section{Global Combining}
\label{sec:combining}
The third and last stage in the global aggregation step is the processing and combining of the received signal $\boldsymbol{y}_t$ at the server into the global model $\myVec{\theta}_t$. 
The goal is to produce an accurate global model, where  accuracy is typically characterized by the global objective \eqref{eq:global_loss}.
We next describe conventional combining strategies.
We then discuss how this operation is viewed from a signal processing perspective, and provide examples of methods for smart combining in light of the mentioned challenges.

\textbf{Conventional Combining:}
FedAvg implements global combining by setting the global model to be a weighted average of the local updates. It is typically assumed that the model updates 
can be accurately recovered from the channel output $\myVec{y}_t$, such that the server has direct access to the model updates of the users participating in the current round, i.e., $\{\myVec{\theta}_t^i- \myVec{\theta}_{t-\SDGIter}^i\}_{i \in \mySet{G}_t}$. The global model is then obtained by
\begin{equation}
\label{eqn:combining1}
    \myVec{\theta}_t = \Phi(\myVec{y}_t) = \frac{N}{|\mySet{G}_t|}\sum_{i \in \mySet{G}_t} p_i (\myVec{\theta}_t^i- \myVec{\theta}_{t-\SDGIter}^i + \myVec{\theta}_{t-\SDGIter}),
\end{equation}
where  $\mySet{G}_t$ is the set of participating devices, and $\myVec{\theta}_{t-\SDGIter}$ is the global model broadcasted on the previous distribution step. 

The weights $\{p_i\}$ in \eqref{eqn:combining1} are the same as in the global objective \eqref{eq:global_objective}, and are typically determined in accordance with the different dataset sizes of the users. Nonetheless, these weights often do not account for other differences between the users, such as the heterogeneous nature of the local data distributions, and the fact that each user's instantaneous updates is expected to have a different contribution on the learned model. 
 Intrinsically, the centralized model combining in \eqref{eqn:combining1} in fact minimizes the loss of the resulting global model with respect to data drawn from a distribution which equals the weighted average distribution of the local datasets \cite{mohri2019agnostic}. 
 
 To formulate this mathematically, let $\mathcal{Q}_i$ be the distribution associated with the dataset at   user $i$. The model which is optimized based on $F(\cdot)$ in fact minimizes the loss measure taken with respect to data from the averaged distribution $\bar{\mathcal{Q}} \triangleq  \sum_{i=1}^N p_i \mathcal{Q}_i$ \cite{mansour2012multiple}.
This implies that ensuring accuracy in the sense of minimizing the global loss may result in a combined model that is unable to accurately infer samples drawn from the distributions under which it was trained. 
In addition, conventional strategies assume that the information sent by the users is reliable, ignoring possible adversaries that can harm the results. 
Straight-forward weighted aggregation of the users' updates can be arbitrarily skewed by a single Byzantine-faulty user via, e.g., poisoning attacks \cite{fang2020local}. 

\textbf{Signal Processing Perspective:} The channel output $\myVec{y}_t$ is statistically related to the model updates $\{\myVec{\theta}_t^i - \myVec{\theta}_{t-\SDGIter}^i\}_{i \in \mySet{G}_t}$. As discussed in the previous sections, this relationship  possibly includes noise and attenuation due the transmission procedure, as well as distortion due to lossy compression and/or privacy amplification. Consequently, mapping   $\myVec{y}_t$ into a global model $\myVec{\theta}_t$ which yields an accurate inference rule in the sense of some risk function is inherently a {\em statistical estimation} setup. 

Nonetheless, the application of statistical estimation techniques for global combining is challenging due to some of the properties of federated learning setups. First, while estimation methods often rely on prior knowledge of the joint distribution of the observations and the desired quantity, one rarely has a access to a reliable characterization of the statistical relationship between the observations and the desired global model. In fact, it is quite unlikely to assume that the server knows the distribution of the model updates. Furthermore, the global objective in  \eqref{eq:global_loss} relies on data and is computed using the datasets available at the users, implying that the server cannot even evaluate the global objective for a given model, let alone optimize its combining rule based on it. Moreover, as discussed earlier, even when the server can characterize a combining rule which yields model that minimizes global objective \eqref{eq:global_loss}, the resulting model may in fact be inaccurate to a large family of relevant distributions. Finally, the server should also be able to cope with unreliable users; a requirement which is not typically encountered in statistical estimation. 

This motivates the design of dedicated global methods oriented to the federated learning framework. In the following, we first discuss schemes which aim at forming accurate global models from the channel outputs in a non-secure manner, i.e., without have to handle unreliable and malicious users.
Then, we review methods for adapting combining methods to be Byzantine-robust, i.e., allow defending against malicious users. 

%
%
\subsection{Federated Combining}
\label{ssec:clustered}
The purpose of the combining stage is to integrate the individual users' updated models into one rich global model, thus gradually refining the inference rule as training progresses. This procedure handles the channel effects, such as the introduction of noise and fading, and aims at having the updated global model yielding an accurate inference rule. 

The direct formulation of this problem uses the global objective \eqref{eq:global_loss}, aiming to set $\myVec{\theta}_t = \Phi(\myVec{y}_t)$ in order to minimize $F( \Phi(\myVec{y}_t))$. As discussed above, when the local datasets are heterogeneous, the model which minimizes $F(\cdot)$ may perform poorly on data drawn from some (or even all) the  distributions used in its training. The effect of statistical heterogeneity can be mitigated by modifying the optimization procedure, effectively resulting in the local models corresponding to a distribution other than the local dataset distribution $\mathcal{Q}_i$. This can be achieved by, e.g., combining additive local and global correction terms to the local optimization
\ifFullVersion
procedure as in \cite{karimireddy2020scaffold}, see also 
\else
procedure, see, e.g., 
\fi
\cite[Sec. 3.2]{kairouz2019advances}. 
Nonetheless, even when the local models are trained using conventional optimizers, which is often the case in federated learning, the harmful effects of statistical heterogeneity can be reduced upon the combining of the model updates, as we show in the following. 

\begin{figure*}[!t]
\centering
\subfigure[Clustered federated learning illustration.]{ 
\label{fig:Clustered1} 
\includegraphics[width=0.35\linewidth]{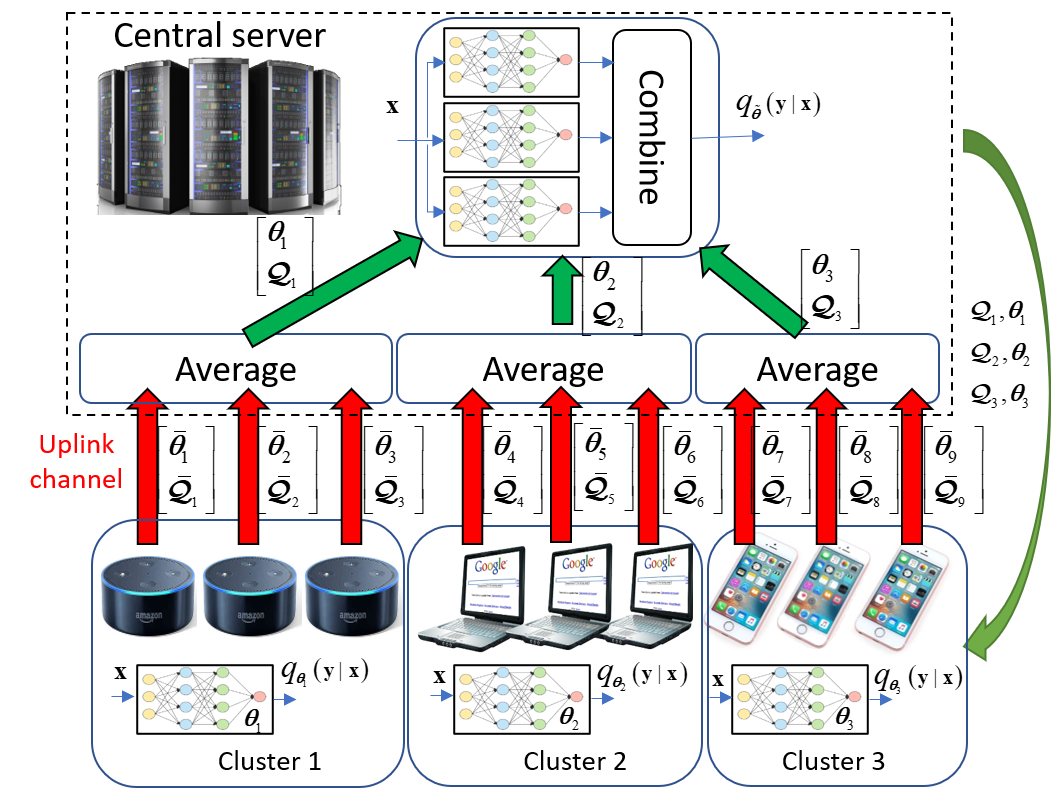}}
\quad
\subfigure[Loss versus model parameters.]{ 
\label{fig:Clustered2} 
\includegraphics[width=0.35\linewidth]{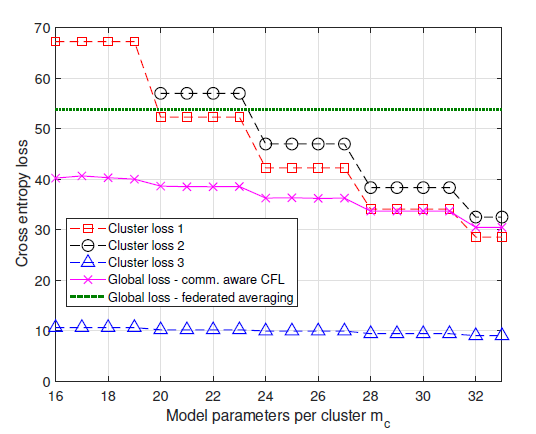}} \caption{\label{fig:ClusteredFL} Clustered federated learning with combined inference rules. The numerical study in (b), adopted from \cite{shlezinger2020communication}, considers the training of linear regression model over data distributed among $N=9$ users, divided into $C=3$ clusters.
In each cluster, the input is randomized in an i.i.d. fashion from a uniform mixture of three identity covariance Gaussian functions.
The results compares the accuracy of the combined global model to conventional federated learning based on FedAvg combining, when both are tested using data taken from a mixture of the cluster distributions, as well that of the individual cluster models tested on data from their local distribution.}
 \vspace{-0.3cm}
\end{figure*}

\textbf{Design Considerations}: 
The design of federated combining methods should account for the following considerations:
\begin{enumerate}
\item Computing the global loss  requires access to the local datasets. In addition, the model updates themselves are determined by the datasets. As these sets are not available to the server, any optimization carried out using either the global loss measures or a statistical characterization of the model updates inevitably requires additional information exchange between the users and the server. This should be carried out in a manner which does not induce notable communication burden as well as does not leak information on the individual  data samples. Alternatively, one can utilize server-side data, when such is available, for optimizing the combining rule.
\item While the edge devices may vary considerably, it is reasonable to assume that devices of similar technology and/or geographic location observe samples of a similar distribution. Tackling heterogeneity in aggregation is notably facilitated if the users can be clustered into groups with similar data distribution, as illustrated in Fig.~\ref{fig:Clustered1}. For instance, such knowledge can be used to determine the combining rule by treating model updates from different clusters differently. Several methods have been recently proposed to enable clustered federated learning, see, e.g., \cite{sattler2020clustered, ghosh2019robust}.  
\end{enumerate}

\textbf{Federated Combining Methods}:
 Broadly speaking, there are two main strategies to aggregate the channel outputs into a global model yielding an accurate inference rule: The first is to combine the model updates into a single model $\myVec{\theta}_t$, while using a mapping $\Phi(\cdot)$ which compensates for channel effects, distortion, and/or bias due to heterogeneity; The second strategy aggregates the model updates into a set of global models, i.e., a mixture of models (also referred to as {\em model interpolation} \cite{mansour2020three}). During inference, the mappings of these models, rather than their parameters, are combined, as illustrated in Fig.~\ref{fig:Clustered1}. 

\begin{enumerate}
    \item {\em Single Combined Model}: As discussed in the previous section, the conventional FedAvg combining rule reduces some of the harmful effects of the distortion and noise induced in the encoding procedure and the communication channel, respectively. However, one can further mitigate these effects by properly accounting for their presence. For instance, the work \cite{lee2020bayesian} proposed to compensate for the distortion induced by coarse quantization by designing the combining rule to approach the minimal mean-squared error estimate of some desirable parameters $\myVec{\theta}_t^{\rm opt}$, i.e., $\Phi(\myVec{y}_t) \approx \mathbb{E}[\myVec{\theta}_t^{\rm opt}|\myVec{y}_t]$. In particular, $\myVec{\theta}_t^{\rm opt}$ was set to the FedAvg combining rule in \cite{lee2020bayesian}, while the conditional expectation was computed based on knowledge of the communication channel and the quantization scheme, as well as additional information conveyed from the users regarding the statistical moments of the model updates.  
    
    Deviating from the combining rule in \eqref{eqn:combining1} can help not only in tackling distortion and noise induced in the encoding and transmission processes, but also to mitigate the harmful effects of statistical heterogeneity. This can be achieved via weighted averaging of the weights as in conventional FedAvg, i.e., $\myVec{\theta}_t = \sum_{i \in \mySet{G}_t} p_i \myVec{\theta}_t^i$ (for full user participation $|\mySet{G}_t| = N$), while optimizing the weights $\{p_i\}$. For instance, the agnostic federated learning scheme of \cite{mohri2019agnostic} learns this combination in a manner which minimizes the maximal loss over a set of weighting coefficients $\mySet{P}$, i.e., the loss measure becomes $\max_{\tilde{p}_i \in \mySet{P}} \sum_i \tilde{p}_i f(\myVec{\theta},\mySet{D}^i)$. The main drawback in learning the combining weights $\{p_i\}$ along with the model parameters $\myVec{\theta}$ stems from the fact that it involves an increased amount of communications between users and the server. 
    
    \item {\em Mixture of Models}: An alternative approach to deal with the harmful effects of statistical heterogeneity in aggregation is to combine the {\em inference rules} rather than the model parameters. Here, letting $q_{\myVec{\theta}}(\cdot)$ denote the inference rule parameterized by $\myVec{\theta}$,  the global inference rule is given by a combination of $\{q_{\myVec{\theta}_t^i}(\cdot)\}$. Using a combination of inference rules rather than a single aggregated model allows utilizing tools from multi-source adaptation, ensemble learning, and mixture-of-experts, to determine how to combine the models. The main challenge associated with this strategy is that a separate model has to be maintained and trained for each user/cluster of users. This increases the volume of the resulting global model, which is comprised of an ensemble of individual models rather than a single aggregated one, and implies that the model distributing step over the downlink should be carried out in a multicast fashion, as different models are distributed to different users, rather than via conventional broadcasting. An example of a mechanism for combining the inference rules rather than the model parameters is detailed as  {\em Heterogeneous Federated Learning via Mixture of Models} on Page~\pageref{Box:Cluster}.
    
\end{enumerate}
Aggregating the parameters into a single model allows all the users to train the same model. Combining the inference rules is oriented towards scenarios in which inference is carried out at the server side, which has access to all the models. Nonetheless, one can still combine diverse inference rules on the edge via collaboration,   as proposed in \cite{shlezinger2021collaborative},  at the cost of additional communications during inference.

\begin{tcolorbox}[float*=t,
    width=2\linewidth,
	toprule = 0mm,
	bottomrule = 0mm,
	leftrule = 0mm,
	rightrule = 0mm,
	arc = 0mm,
	colframe = myblue,
	colback = mypurple,
	fonttitle = \sffamily\bfseries\large,
	title = Heterogeneous Federated Learning via Mixture of Models]
    Consider a federated learning system with $N$ users, where the data at the $i$th user follows a marginal distribution, denoted $\mathcal{Q}_i(\cdot)$. Each user uses its local dataset $\mySet{D}^i$ to train a local model $\myVec{\theta}_t^i$ which dictates an inference rule $q_{\myVec{\theta}_t^i}(\myVec{a})$ when applied to sample $\myVec{a}$. 
    The users are divided into $C\leq N$ clusters (preferably $C \ll N$), where each cluster is characterized by a single input distribution measure $\mathcal{Q}_c(\cdot)$.
    
    In combining via mixture of models, the server orchestrates the training of a separate model for each cluster via conventional federated averaging. In such cases, $C$ separate global models are simultaneously trained in a federated manner \cite{ghosh2019robust} using (approximately) homogeneous datasets.  By letting $\tilde{\myVec{\theta}}_t^c$ be the global model parameters of the $c$th cluster, the global inference rule is given by $  q_{\myVec{\theta}}(\myVec{a}) = \sum \limits_{c=1}^C \alpha_c q_{\tilde{\boldsymbol{\theta}}_c}(\myVec{a})$, where $\{\alpha_c\}$ are the averaging coefficients. 
    
    The work \cite{shlezinger2020communication} proposed to use averaging coefficients which are parametric estimates of the cluster marginals, i.e., each $\alpha_c$ is a function of the input $\myVec{a}$ such that $\alpha_c(\myVec{a}) \approx \frac{\mathcal{Q}_c(\myVec{a})}{\sum_{k=1}^C \mathcal{Q}_k(\myVec{a})}$. This implies that an additional model for estimating the marginal distribution is trained for each cluster, at the cost of additional communication overhead.
    An illustration of this procedure with $N = 9$ users and $C = 3$ clusters is depicted in Fig. \ref{fig:Clustered1}.
    
    Such aggregation can be shown to yield reliable inference when applied to samples from different distributions, which are within some proximity of the cluster marginals $\{\mathcal{Q}_c\}$. In particular, if the loss measure of the model of cluster $c$ is upper bounded by $\epsilon_c$, then for samples drawn from any distribution which is $\rho$-norm bounded (see \cite{mansour2012multiple}) by $\{\mathcal{Q}_c\}$, the expected loss can be upper bounded by $\frac{1+\delta}{1-\delta}\rho \sum_{c=1}^C \epsilon_c$, where $\delta>0$ is a bound on difference between the estimated cluster marginals and the true ones (see \cite{shlezinger2020communication} for the exact statement of the difference).
    This indicates that one can achieve arbitrarily small global loss for the individual distributions observed at each cluster in "test" time.
    Another insight arising from the result is that the individual cluster loss is directly related to the richness of its model. 
    This is also illustrated in Fig. \ref{fig:Clustered2}, which also compares to combining based on FedAvg.
    
\label{Box:Cluster}
\end{tcolorbox}

%
%
\subsection{Security-Enhanced Federated Combining} 
\label{ssec:truncation}
The fact that many different users take part in  federated learning  gives rise to security issues which are not commonly encountered in conventional centralized learning. 
In particular, model updates sent by some users can be unreliable and significantly deviate from their normal values, either unintentionally or maliciously. This in turn induces bias on the learned model, and may severely degrade the convergence performance.

The need to be able to cope with unreliable users falls under the framework of distributed robust learning \cite{chen2017distributed}, also referred to as {\em Byzantine fault tolerant} or {\em Byzantine robust} learning.
Distributed robust learning is concerned with samples containing a mixture of authentic samples and outliers held by a central server. The authentic samples are generated according to an underlying model (i.e., ground truth), and the outliers may be arbitrarily corrupted or even maliciously chosen. 
Such robust aggregation methods are typically non-affine, and are designed not to be affected by the values of the extreme (and possibly unreliable) samples. 
A common way to reduce the effect of the malicious outliers on the learned model is to use the geometric median in the aggregation procedure instead of the simple averaging. The geometric median is a generalization of median in one-dimension to multiple dimensions. It aggregates a collection of independent estimates into a single estimate with significantly stronger concentration properties, even in presence of a constant fraction of outliers in the collection.

\textbf{Design Considerations}: 
For a federated learning aggregation method to be Byzantine fault tolerant, it must not be affected by abnormal model updates, which  correspond to unreliable users.
A key challenge which arises in Byzantine-robust federated combining stems from the fact that the aggregation technique should not only mitigate the effect of abnormal updates values, but also do so in a way that does not significantly impair the optimization performance, and without inducing additional significant communications between the users and the server.
Thus, an important question in this setting is what levels of accuracy should one expect when training a model in a federated manner while being Byzantine-robust, and how to design algorithms that improve the accuracy. 

In particular, the design of Byzantine-robust federated combining methods must account for the following consideration:
\begin{enumerate}
    \item  The heterogeneous nature of the users must be accounted for when boosting Byzantine fault tolerance. Specifically, not only are the users outside the control of the server, but also data may not be homogeneous between users, and different users possess different amount of data. Thus, a relatively large level of diversity, which is typically used in Byzantine-robust distributed optimization as measure of abnormality, is in fact expected in federated learning, even when all the users are  reliable.
    \item Unreliable users may not only provide corrupted model updates, but also affect the aggregation rule. For instance, conventional FedAvg weights the updates based on the number of samples each user has, since $p_i$ in \eqref{eq:FedAvg_Server} is usually set to $p_i = \frac{n_i}{\sum_j n_j}$. Unreliable users may  thus report a false dataset size to modify the aggregation rule. This ability of Byzantine-faulty entities to corrupt not only their own contribution, but also how it is processed by the server, is a unique challenge arising in federated learning. 
    \item In general, Byzantine-robust combining methods require knowledge on the maximal fraction of unreliable users. Nonetheless, some methods are invariant to this requirement, as discussed in the sequel. 
\end{enumerate}

\textbf{Robust Combining Methods}:
 FedAvg combining is ignorant of the value of the model updates, and is thus sensitive to corrupted models. Therefore, Byzantine robust combining methods must account for the specific values of the updates, and provide  means for identifying which of the updates are treated as abnormal.
The leading methods to realize robust aggregation are based on replacing the conventional averaging with one of the following non-affine computations:
\begin{enumerate}
    \item {\em Median} aggregation is highly robust to abnormal updates. For instance, the one dimensional median operation has the property that if over half of the samples points lie in some range $[-r,r]$ for a given $r >0$, then the median must be in $[-r,r]$. Likewise, in multiple dimensions, the geometric median has a similar robust property \cite{chen2017distributed,yin2018byzantine}. The method of geometric median can be further generalized to marginal median and to “mean around median” \cite{xie2018generalized}.  When used for aggregating the model updates, median-based aggregation guarantees that the difference in the global model is in the proximity of at least half the considered updates, which yields resiliency to abnormal updates. Such robust aggregation does not require knowledge on the fraction of unreliable users.  
    \item {\em Krum} aggregation bears some similarity to median aggregation in the sense that it selects a single model update to use for updating the global model. Here, the model update is selected as the one which is the  closest to a given set of neighbouring model updates.  
    \item {\em Truncation} mappings first discards a subset of the model updates, which are treated as abnormal, and then aggregates the remaining updates via FedAvg \cite{yin2018byzantine}. 
\end{enumerate}
The above methods are designed to constitute a robust alternative to mere averaging. They can also be modified to replace weighted averaging, e.g.,  when the datasets are imbalanced, by applying them to the weighted model updates scaled by their proportional coefficient $N\cdot p_i$. Furthermore, the challenge with dealing with diversity due to heterogeneity in a Byzantine robust manner can be tackled by dividing the users into clusters, and aggregating in a robust manner in each cluster separately \cite{ghosh2019robust}. 
%
An example of Byzantine-robust federated learning methods using the median and truncation combining strategies is given on Page \pageref{Box:Bizantine}.

\begin{figure*}[!t]
\centering
\subfigure{ 
\label{fig:Malicious1} 
\includegraphics[width=0.42\linewidth]{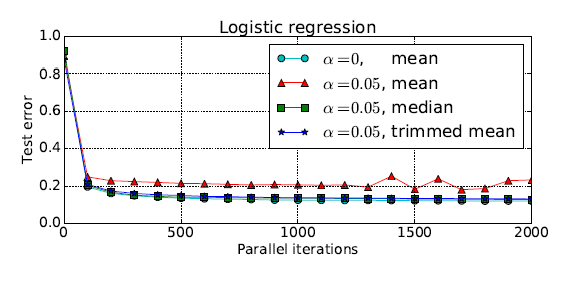}}
\quad
\subfigure{ 
\label{fig:Malicious2} 
\includegraphics[width=0.42\linewidth]{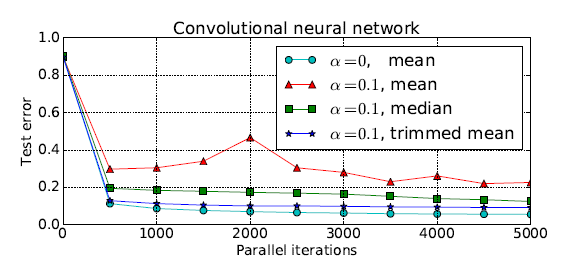}} 
 \caption{\label{fig:Malicious} Numerical study of the coordinate-wise median and trimmed-mean methods adopted from \cite{yin2018byzantine} with the authors' permission. 
 The scenario considered is handwritten digit classification based on the MNIST dataset distributed in an i.i.d. fashion between $N =40$ users in the logistic regression model, and $N =10$ in the convolutional neural network model. The methods are compared to the vanilla distributed gradient descent (aggregating the gradients by taking the mean), where the trimmed mean uses $\beta = \alpha$.}
 \vspace{-0.3cm}
\end{figure*}

\begin{tcolorbox}[float*=t,
    width=2\linewidth,
	toprule = 0mm,
	bottomrule = 0mm,
	leftrule = 0mm,
	rightrule = 0mm,
	arc = 0mm,
	colframe = myblue,
	colback = mypurple,
	fonttitle = \sffamily\bfseries\large,
	title = Byzantine-Robust Federated Learning]	
    Consider a federated learning system with $N$ users possessing datasets of identical sizes, i.e., $|\mySet{D}^i| \equiv n$.
   Let $\alpha$ denote the fraction of the $N$ users are unreliable, and the remaining $1-\alpha$ fraction are normal, and denote the model updates by $\boldsymbol{g}_t^i = [g_{t,1}^{i}, \ldots g_{t,d}^{i}] \triangleq \myVec{\theta}_{t}^i- \myVec{\theta}_{t-\SDGIter}^i$.
    At each communication round, the model updates are sent to the server for aggregation, where the Byzantine users can send arbitrary messages; in particular, they may have complete knowledge of the system and learning algorithms, and can collude with each other.
    Assume an orthogonal transmission over noiseless channel, and let $\boldsymbol{y}_t$ be the channel output contains the received model updates, i.e.,  $\boldsymbol{y}_t = [\boldsymbol{g}_t^1,\ldots,\boldsymbol{g}_t^N]^T$. 
    
   Two combining strategies proposed in \cite{yin2018byzantine} to address malicious users are coordinate-wise median and coordinate-wise trimmed mean. 
   The coordinate-wise median method aggregates the local updates using one dimensional median: 
   \begin{equation}
   \label{eq:median}
    \boldsymbol{g}_t(\boldsymbol{\theta}) = \text{med} \{\boldsymbol{y}_t\},
    \end{equation}
    where $\boldsymbol{g}_t$ is a vector with its $k$ coordinate being $g_{t,k} = \text{med} \{g_{t,k}^{i}: i\in \mathcal{N} \}$ for each $k \in \{1 \ldots d\}$.   
    
    The coordinate-wise $\beta$-trimmed mean method removes the largest and smallest $\beta$ fraction of the $k$th coordinate of the model update vector, i.e.,: 
    \begin{equation}
    \label{eq:trimmed}
        g_{t,k} = \text{trmean}_\beta \{\boldsymbol{y}_{t,k}\} = \frac{1}{(1-2\beta)N} \sum_{g \in U_k} g,
    \end{equation}
    where $U_k$ is a subset of $\{g_{t,k}^1, \ldots g_{t,k}^N \}$ obtained by removing the largest and smallest $\beta$ fraction of its elements. Typically, $\beta$ is set to be not smaller than the expected portion of unreliable users, which is denoted by $\alpha$.

The two strategies are designed in order to reduce the affects of possible outlires before the global aggregation. 
Both aggregation strategies $(\ref{eq:median})$ and $(\ref{eq:trimmed})$ have provable guarantees on the convergence rates.

The work \cite{yin2018byzantine} characterized the accuracy of such Byzantine-robust aggregation, focusing on setups where each user sends the loss gradients, i.e., $\SDGIter =1$, and the loss surface is strongly-convex smooth and with bounded skewness. It was shown in  \cite{yin2018byzantine}  that in such cases, the  expected  gap  between  the  achieved  loss  function using the coordinate-wise median method and the minimum loss is an increasing function of $\alpha$, and a decreasing function of $n$ (specifically with rate $\frac{1}{n}$).
For strongly-convex smooth objectives and sub-exponential gradients, and by setting  $\beta \geq \alpha$, the  expected  gap  between  the  achieved  loss  function using the coordinate-wise $\beta$-trimmed mean method and the minimum loss is an increasing function of $\alpha$, and a decreasing function of $n$ (specifically with rate $\frac{1}{\sqrt{n}}$).
Although the median-based algorithm has a poorer convergence rate with respect to the parameter $n$, it requires milder tail/moments assumptions, and does not requires the knowledge of the maximal number of expected unreliable users, encapsulated in the portion $\alpha$.
The performance of the two methods under different values of $\alpha$ and loss functions are shown in Fig.~\ref{fig:Malicious}. 

\label{Box:Bizantine}
\end{tcolorbox}	

\section{Future Research Directions}
\label{sec:future_research}
We conclude the article with a discussion on some of the future research directions, which can further strengthen the role of signal processing and communications in the emerging federated learning paradigm. We follow the division of the global aggregation step used in the article; discussing future research directions related to the users' local updates encoding, after which we elaborate on those focusing on the transmission of the encoded model updates and their combining on the server side. Finally, we discuss additional future directions which go beyond this division. 

\subsection{Local Updates Encoding}
\label{ssec:future_encoding}
Processing of the updated models produced by the local optimization procedure allows to tackle some of the core challenges of federated learning. As discussed in the ``\nameref{sec:encoding}" section, by encoding the model updates prior to their transmission, one can notably relieve the communication burden via, e.g., quantization and compression, as well as boost privacy preservation  using encryption and differential privacy techniques.

The potential of learning-aware local encoding  in facilitating federated learning at a large scale gives rise to a multitude of possible research directions. Most studies to date on uplink compression focus on a single user, assuming that all users in the network utilize the same encoding mechanism. Nonetheless, the heterogeneous nature of federated learning motivates treating the need to compress the model updates from a perspective of the overall distributed system. For instance, one would expect users of different technologies to utilize different quantization resolution, as for each device it is expected to be translated into communication delay in a different manner. Furthermore, it may be preferable to change the compression mapping as the learning procedure progresses, using different resolutions  at distinct communication rounds, aiming to minimize the overall communication burden. Finally, for users with datasets of similar distributions, the model updates may  be statistically correlated, indicating the possibility to exploit this correlation to further reduce the communication load via, e.g., distributed source coding techniques. 

Furthermore, the fact that model updates are often compressed in a manner which induces distortion motivates the analysis of lossy compression as a privacy preserving mechanism. In particular, differential privacy amplification methods rely on adding artificial distortion to the model updates, where the motivation is to obscure them such that each single data sample used in their training cannot be recovered. Since quantization and lossy compression already induces some level of distortion, one can  reduce the level of artificial noise needed to guarantee differential privacy. Furthermore, the fact that quantization discretizes the model updates facilitates applying encryption methods, which often assume integer quantities.

\subsection{Uplink Transmission}
\label{ssec:future_Transmission}
The main challenge in uplink transmission stems from the fact that communications is often carried out over shared noisy channels which are limited in their spectral and temporal resources. As discussed in the ``\nameref{sec:Transmission}" section, the delay induced by communications and its harmful effect on the convergence time and on the accuracy of the model learned in a federated manner can be significantly reduced by dividing the channel resources in learning-aware fashion. For instance, by boosting interference-free orthogonal communications while prioritizing users which are expected to have a greater contribution to the learning process,  and even by promoting full reuse via over-the-air computation.  

Nonetheless, there is still much to be explored in the study of uplink transmission schemes for federated learning. In particular, while user selection methods give all users the chance to participate, the prioritization among the users should account for a multitude of considerations, which are not accounted for to date. For instance, the fact that edge devices are comprised of a broad range of different technologies indicates that some devices are more likely to transmit at increased delays. Statistical heterogeneity also plays an important role, particularly when one has prior knowledge on the distribution of each user as in clustered federated learning \cite{sattler2020clustered}; Here, one should consider whether it is preferable to assign resources and select users in a diverse or homogeneous fashion. 

Furthermore, resource allocation and user selection can also contribute to non-orthogonal transmissions. Here, it is possible to rely on dynamic spectrum access schemes, commonly studied for wireless sensor networks. Developing dynamic spectrum access  strategies to federating learning tasks is subject to challenges which are not encountered in its traditional domains. For instance, intriguing research questions are: (i) How to develop dynamic spectrum access that integrates over-the-air and orthogonal transmissions when coherent over-the-air transmission over the entire network is not efficient; and (ii) How to develop spatio-temporal dynamic spectrum access strategies when users experience deep fading in time and frequency. For example, a promising direction is to model the fading channel as a known or unknown Markovian process, for which theoretical performance measures for channel allocation strategies can be developed rigorously as in, e.g.,  \cite{gafni2021distributed}.

Finally, we note that the vast majority of the existing federated learning algorithms focus on communications carried out over networks obeying a star topology, in which all users communicate directly with the server. Nevertheless, common communication schemes in wireless networks  allow multi-hop communications between edge devices, either via wireless access points, relays, or even using the emerging technology of reconfigurable intelligent surfaces. In such cases, different subsets of the users may share distinct channels, while communications on both the uplink and the downlink is carried out over a multi-hop route.  Learning in a federated manner over hierarchical communication networks requires a dedicated treatment regarding the division of channel resources, the presence of intermediate aggregations, and the possibilities of model caching along the routes as well as of carrying out over-the-air federated learning in a hierarchical manner.  

\subsection{Global Combining}
\label{ssec:future_Combining}
Global combining refers to how the channel output observed by the server orchestrating the federated learning procedure is translated into an updated global model. As discussed in the ``\nameref{sec:combining}" section, applying combining mechanisms other than the conventional federated averaging rule \eqref{eq:FedAvg_Server} allows to compensate for distortions induced in encoding and transmission of the model updates, mitigate the harmful effect of unreliable and malicious users, and facilitate accurate inference in the presence of heterogeneous data division. 

The fact that most non-secure combining mechanisms for tackling heterogeneity rely on weighted averaging motivates exploring non-linear combinations. Such an analysis can be combined with Byzantine-robust aggregation, which inherently utilizes non-linear mappings to omit outliers, exploring the joint functionality in training a model capable of accurately inferring using each of the individual distributions used for its training, possibly yielding personalized models \cite{mansour2020three}, while being tolerant to the presence of malicious users. 

Furthermore, the heterogeneous nature of the data and the users in federated learning systems implies that mechanisms for truncating the contribution of unreliable users must account for such properties. One approach to do so is to divide the users into clusters based on their distribution and implement Byzantine-robust aggregation separately over each of these homogeneous clusters, as proposed in \cite{ghosh2019robust}. The fact that such a strategy already trains multiple models for each cluster motivates its combination with aggregation mechanisms which combine the inference rules during test. 

%
%
\subsection{Additional Directions}
\label{ssec:future_Additional}
Our description so far follows the division of the global aggregation stage in the federated learning flow into the aforementioned encoding, transmission, and combining stages. This structured framework enables identifying the relationships between federated learning schemes derived for different purposes, as well as facilitates the derivation of dedicated future methods. Nonetheless, one is not limited to design mechanisms within this framework, and it is natural to also consider joint optimization of multiple steps. Furthermore, while most of the key challenges of federated learning which are natural to address using signal processing techniques lie in the global aggregation stage, the federated learning flow is also comprised of the model distributing and local training steps. It is thus of interest to extend the scope beyond the aggregation of the model updates into a global model.

One such research direction involves the joint optimization of the multiple stages of which global aggregation is comprised. For instance, jointly treating the local encoding and uplink transmission stages can be studied as a distributed joint source channel coding setup, possibly minimizing the delay in conveying the overall model updates. In addition, it is likely that one can further benefit in terms of convergence speed by designing such direct mapping from model updates into channel input of the complete set of users in a joint manner with the combining mechanism at the server side.   

Additional relevant research directions which go beyond the division considered in this article focus on reducing the communication overload and boosting privacy in the presence of external adversaries during downlink transmission. Another possible direction is to improve the convergence time in the local training stage without modifying the optimization algorithm by properly selecting the data. Here, one can consider choosing a subset of the data used for training via active learning techniques, which dates back to Chernoff’s framework of optimal experimental design \cite{chernoff1959sequential}, in order to reduce the training time and thus the overall convergence delay. An intriguing question is thus how to select the useful data among multiple users in the heterogeneous federated learning setting. This will allow significant bandwidth and energy savings by prioritizing informative data sources. 

Finally, while federated learning deals with the training of a machine learning model, its distributed nature also motivates the study of how to use this learned model during inference. For instance, federated learning carried out in a clustered fashion \cite{ghosh2019robust,shlezinger2020communication} results in different users having access to different diverse models, as well as a centralized server which has access to all these models. In this case, one can consider scenarios in which multiple edge users collaborate in inference as a form of wisdom of the crowds as proposed in \cite{shlezinger2021collaborative}, or alternatively, combining server-based inference with localized one to enabled personalized decision making.

\bibliographystyle{IEEEtran}

\end{document}